\documentclass{article}

\PassOptionsToPackage{numbers, compress}{natbib}
\usepackage[preprint]{neurips_2020}

\usepackage[utf8]{inputenc} % allow utf-8 input
\usepackage[T1]{fontenc}    % use 8-bit T1 fonts
\usepackage{hyperref}       % hyperlinks
\usepackage{url}            % simple URL typesetting
\usepackage{booktabs}       % professional-quality tables
\usepackage{amsfonts}       % blackboard math symbols
\usepackage{nicefrac}       % compact symbols for 1/2, etc.
\usepackage{microtype}      % microtypography

\usepackage{mathtools}
\usepackage{multirow}
\usepackage[table,xcdraw]{xcolor}
\usepackage{floatrow}
\usepackage{amsthm}
\usepackage{subfigure}
\usepackage{url}
\usepackage{enumitem}
\newcommand*{\p}{\mathbb{P}}
\newcommand*{\Beta}{\mathcal{B}}
\newcommand{\xhdr}[1]{\vspace{0.1mm}\noindent{{\bf #1.}}}
\newfloatcommand{capbtabbox}{table}[][\FBwidth]
\newtheorem{proposition}{Proposition}[section]

\title{A Bayesian Hierarchical Network for Combining Heterogeneous Data Sources in Medical Diagnoses}

\author{%
  Claire Donnat \\ 
  Department of Statistics\\
Stanford University\\
  \texttt{cdonnat@stanford.edu} 
  \And 
  Nina Miolane \\ 
  Department of Statistics\\
  Stanford University\\
    \texttt{nmiolane@stanford.edu} \\
   \And
  Jack Kreindler \\
  Centre for Health and Human Performance\\
  \texttt{drjack@chhp.com}
   \And 
  Frederick de Saint Pierre Bunbury\\
  Carnegie Institution for Science\\
  \texttt{fbunbury@carnegiescience.edu}
\\
}

\begin{document}

\maketitle

\begin{abstract}

Computer-Aided Diagnosis has shown stellar performance in providing accurate medical diagnoses across multiple testing modalities (medical images, electrophysiological signals, etc.). While this field has typically focused on fully harvesting the signal provided by a single (and generally extremely reliable) modality, fewer efforts have utilized imprecise data lacking reliable ground truth labels. In this unsupervised, noisy setting, the robustification and quantification of the diagnosis uncertainty become paramount, thus posing a new challenge: how can we combine multiple sources of information -- often themselves with vastly varying levels of precision and uncertainty -- to provide a diagnosis estimate with confidence bounds? Motivated by a concrete application in antibody testing, we devise a Stochastic Expectation-Maximization algorithm that allows the principled integration of heterogeneous, and potentially unreliable, data types. Our Bayesian formalism is essential in (a) flexibly combining these heterogeneous data sources and their corresponding levels of uncertainty, (b) quantifying the degree of confidence associated with a given diagnostic, and (c) dealing with the missing values that typically plague medical data.
We quantify the potential of this approach on simulated data, and showcase its practicality by deploying it on a real COVID-19 immunity study.
\end{abstract}
\vspace{-0.3cm}

\section{Introduction and Related Work}
\vspace{-0.3cm}
% Traditional diagnosis and ML

%Clinicians traditionally perform a diagnosis by integrating different sources of information, typically combining
Current medical diagnoses are most often based on the combination of several data inputs by medical experts, typically including (i) clinical history, interviews, and physical exams, (ii) laboratory tests, (iii) electrophysiological signals, and medical images. Advances in the machine learning (ML) community have highlighted the potential of ML to contribute to the field of Computer-Aided Diagnosis (CAD), for which we distinguish two main classes of methods.

% ML contributions: one data sources

\xhdr{Single modality analysis} Most recent efforts in the ML community have focused on analyzing a single medical data source --- called a ``modality". For instance, the parsing of electronic health records (EHR) for clinical history, patients' interviews, and physical exams has shown huge potential for the diagnosis of a broad range of diseases, from coronary artery disease to rheumatoid arthritis \cite{Harrell1984RegressionPrediction, Kurt2008ComparingDisease, Carroll2011NaiveArthritis., Hippisley-Cox2013PredictingScore, Rahimian2018PredictingRecords}. ML algorithms have been developed to process various types of laboratory results, from urine steroid profiles to gene expression data \cite{Huang2018NovelClassification, Wilkes2018UsingProfiles, Demirci2016ArtificialAlgorithms, Agrahari2018ApplicationsMalignancies}. Others have shown success in automatically processing and classifying medical signals such as electrocardiograms (ECG) \cite{Acharya2017AutomatedNetwork, Kiranyaz2016Real-TimeNetworks, Rad2017ECG-BasedAnalysis, Pawiak2018NovelSystem}, or electroencephalograms (EEG) \cite{Richhariya2018EEGMachine}. Finally, a large body of work has focused on medical imaging analysis encompassing a vast number of tasks, such as automatic extraction of diagnostic features \cite{Thomaz2017FeatureImages, Bar2018ChestTraining}, segmentation of anatomical structures \cite{Moeskops2016AutomaticNetwork, Li2015AutomaticNetworks, Shelhamer2017FullySegmentation, Milletari2016V-Net:Segmentation, Dolz20183DStudy, Christ1919AutomaticFields}, or direct diagnosis through image classification \cite{Anthimopoulos2016LungNetwork, Miki2017ClassificationNetwork}. 

While these works achieve record-breaking diagnostic accuracy, they often rely on supervised learning approaches -- requiring the diagnosis ground-truth to be available during training -- and on the acquisition of large datasets. Furthermore, they focus solely on a single type of data input (EEGs, scans, etc.) -- often acquired by clinicians using specialized, highly accurate equipment -- and do not harvest the potentially rich and complementary sources of information provided by alternative medical modalities. Yet, with the development of at-home diagnostic tests (lateral flow assays, questionnaire data for disease screening, mobile health apps, etc.), this paradigm shifts, and diagnoses have to be established through the combination of multiple sources of cheaper, yet often noisier and more imprecise data. 
%The fusion of these different data sources becomes crucial in providing a more accurate diagnosis. 
In light of the uncertainty exhibited by these various inputs, it also becomes indispensable to pair the provision of a diagnosis with a notion of a confidence interval, thereby thoroughly characterizing our state of knowledge given the available data. 
%--especially in the presence of unreliable tests, which are known to have low sensitivity (false negatives) or specificity (false positives).

% Fusion of different data sources

\xhdr{Multiple modality integration} Interestingly, diagnosis studies fusing different data sources seemed to be more prominent in the first years of CAD. Bayesian networks \cite{Neal1993ProbabilisticMethods} -- also called belief networks -- have been used as a crucial decision tool for automatic diagnosis. Such networks provide a biologically-grounded and interpretable statistical framework that integrates the probabilities of contracting the disease given a patient's clinical history, and the probability of developing symptoms or observing positive test results in the presence or absence of the disease. The advantages of these Bayesian networks are that they are (a) fully unsupervised and do not require ground-truth information, and (b) able to provide meaningful results even in low sample regimes.

Bayesian networks have thus been implemented in a variety of contexts to integrate clinical data and laboratory results, and diagnose conditions ranging from pyloric stenosis to acute appendicitis \cite{Alvarez2006EvaluationStenosis, Gevaert2006PredictingNetworks, Sakai2007AccuracyModel, Gonzalez-Lopez2016DevelopmentUveitis,Sangamuang2018AccuracyDecision}. They have also been used to combine clinical data with medical images, and subsequently applied to assess venous thromboembolism \cite{Kline1984DerivationThromboembolism}, community-acquired pneumonia \cite{Aronsky1998DiagnosingNetwork.}, head injury prognostics \cite{Sakellaropoulos1999DevelopmentTechniques}, or to predict tumors \cite{Wang1999Computer-assistedNetwork,Kahn1997ConstructionCancer, Sneha2013TowardsImages}. As early as 1994, full integration of clinical data, laboratory results and imaging features was performed to diagnose gallbladder disease \cite{Haddawy1994ADisease}. 
%Yet, the data sources provided as inputs are entered manually by the researchers: the imaging features are extracted by a clinician expert, thus failing to leverage the immense machine learning developments described in the previous subsection. Additionally, while the sensitivity and specificity of the laboratory or imaging test results are taken into account, their heterogeneity among tests is not. 

However, contrary to our proposed setting, the uncertainty that these Bayesian networks integrate is typically known and controlled. The medical conditions that they study are well characterized by a set of specific questions and physiological exams (e.g. projectile vomiting, potassium levels and ultrasounds in the case of Pyloric stenosis). Not only do these inputs provide a very strong signal for the diagnosis, but the uncertainty arising from the different modalities can often be reliably informed by a test manufacturer, extensive medical research or prior ML studies (for diagnostic inputs obtained through classification algorithms). The uncertainty of a diagnostic method, however, is not always well specified: whether it be (i) a physician's assessment of the symptoms associated with a novel or rare disease, (ii) predictions of an ML algorithm whose accuracy has not been fully characterized outside of a curated research environment or reference datasets (such as MNIST, CIFAR), or simply (iii) biological tests whose sensitivity still has to be determined. 

\xhdr{Multiple noisy modality integration: a COVID-19 case study} To motivate this paper and further understand the limitations of these previous approaches, let us consider a particular use case: COVID-19 antibody testing. 
%At the time of writing, 
Lateral flow immunoassay (LFA) antibody tests for COVID-19 are one of the manageable, affordable and easily deployable options to allow at-home testing for large population and provide assessments of our immunity.
%--- a crucial step in the easing of current social restrictions and/or lockdown measures.
%At the time of writing, the optimal strategy for a return to usual social, economic and health care practice seems to require robust and accurate assessments of our immunity as lockdown is eased. This requires us to diagnose the immunity of citizens at scale, through the use of testing kits provided by different manufacturers, with variable sensitivity and specificity. In the example of COVID-19 diagnosis, lateral flow immunoassay (LFA) tests are one of the manageable, affordable and easily deployable options to allow at-home testing for large population. These advantages come at the cost of a somewhat low diagnosis accuracy, compared to the more expensive hospital-based diagnostic devices, such as Enzyme linked immunosorbent assays (ELISAs). 
Yet, studies have shown that the sensitivity of these tests remains highly variable and highly contingent on the time of testing.
%, with an 81.1\% sensitivity for tests taken within 10 days of the illness onset. 
The successful deployment of LFAs thus depends on their augmentation with additional data inputs, such as user-specific risk factors and self-reported symptoms.
%, in order to provide a more reliable estimate of the presence of antibodies. 
%As such, one of the possible solutions to deconvolve false negatives from real ones, and improve diagnosis accuracy, is to enrich the LFA results with additional modalities, such as self-reported user-specific risk factors and exhibited symptoms.
%In this context, the current WHO, and UK MRHA, position is that such tests should only be used in a research setting (WHO, Scientific Brief, April 2020).
%Lateral flow immunoassay for COVID-19 testing is a case in point, and the motivating thread behind our analysis. 
%To overcome the inherent limitations of medical modalities such as LFA, we develop a statistical algorithm allowing the provision of each user with a more reliable diagnostic and associated confidence's quantification -- both based on the upload of an image of the LFA test result and additional self-reported clinical data such as prior symptoms and context information (number of people in the household, proximity with others over the past weeks, etc.). This model builds on research on Bayesian networks as a diagnostic tool, adding however, the variable uncertainty in the accuracy associated with the different testings kits. 
%At the individual level, this allows the disentanglement of false negatives from true ones. At the global scale, 
The provision of confidence scores is essential to flag potential false negative or positive tests (requiring re-testing or closer scrutiny) and to assess local prevalence levels -- both pivotal for researchers and policymakers in the context of a pandemic.

\xhdr{Contributions and outline} This applications paper is geared towards the practical integration of noisy sources of information for CAD. Our contribution is two-fold. From a methods perspective, we account for the variability of the inputs by devising a two-level Bayesian hierarchical model. %adapted to the handling of the uncertainty, % that allows us to combine heterogeneous medical data and to account for the variability of diagnostics' accuracies. 
%Our goal is to learn and/or adapt the accuracy associated with each diagnostic input, in order to output both an estimate of the probability for a positive diagnosis and the associated confidence bounds. 
In contrast to existing Bayesian methods for CAD, our model is deeper, and trained using a Stochastic Expectation Maximization (EM) algorithm \cite{celeux1995stochastic,Celeux2001OnAlgorithm,nielsen2000stochastic}. The Stochastic EM allows to overcome the limitations of its non-stochastic counterpart \cite{celeux1995stochastic}, that is (a) its sensitivity to the starting position, (b) its potential slow convergence rate, and (c) its possible convergence to a saddle position instead of a local maximum. From an applications perspective, we gear this algorithm to enhance at-home LFA testing in the context of the COVID-19 pandemic. In particular, we wish to (a) quantify the benefit of multimodal data integration when the diagnostics are uncertain, and (b) show how our method can benefit medical experts or researchers in real life. 
%Section~\ref{sec:data} details the dataset of interest. 
Section~\ref{sec:multimodal} presents the Bayesian model for multimodal integration and the Stochastic Expectation-Maximization algorithm that performs principled and scalable inference. Section~\ref{sec:simu} presents extensive tests of our model on simulated datasets. Section~\ref{sec:real} details the results obtained on the Covid-19 dataset and the impact of our method for affordable and reliable at-home test kits.

%While our problem is spun from a practical, current question, the statistical and modeling challenges that it raises are thus still areas of active research. Indeed, this is an unsupervised lOur goal is to provide each user with (a) the confirmation of the diagnostic, and (b) confidence intervals associated with the uncertainty to shed more light on the uncertainty associated with the diagnostic; while providing policy makers with (c) aggregated analysis of the herd immunity, with associated measures of uncertainty.

%\vspace{-0.4cm}

%While this work is motivated by LFA tests and the Covid-19 pandemic, we highlight that this provides general framework for multimodal integration of medical data for diagnosis purposes. In particular, this is directly extendable to settings where we have multiple test outputs and want to combine them. In particular, work that try to integrate medecial consults held via video, and questionnaire output could benefit from this approach [REF--- what did Serena end up doing with the videos?]

\section{Covid-19 Dataset}\label{sec:data}
%\vspace{-0.3cm}

By way of clarifying the challenges that our algorithm proposes to overcome, we present here the COVID Clearance Remote Recovery \& Antibody Monitoring Platform study 
\footnote{COV-CLEAR, \url{www.cov-clear.com}}, which motivated our approach. The purpose of this study is to track the evolution of the immunity of a cohort of adult participants in the UK with various COVID-19 exposure risks. This paper focuses on a subset of 117 healthcare workers, for which we have both questionnaire information and LFA test results.
%(i) healthcare workers, (ii) household contacts of healthcare workers, (iii) individuals confirmed to have had COVID19, typically through admission to hospital, (iv) a clinically vulnerable cohort of immuno-compromised / cancer patients, (v) university students. 

\xhdr{Home-testing Immunoassay Data} Participants are issued packs of home-testing kits with written instructions. These kits identify Immunoglobulin M (IgM) and/or Immunoglobulin G (IgG) specific to SARS-CoV2 in blood samples using three gold-standard methods \footnote{Home-sampled (blood) Abbott ARCHITECT® Anti-SARS-CoV-2 IgG CMIA CE Marked; PHE Approved} \footnote{Elecsys® Anti-SARS-CoV-2 Double Antigen, CE Marked, PHE Approved} \footnote{Biopanda Ltd  Product Number: RAPG-COV-019 version 1 In-Vitro LFT IgM IgG Antibody test. CE certified for Professional Use Only.}. Pictures and typical examples of LFA test results are provided in the supplementary materials. For the LFAs, participants self-report a positive result if IgM and/or IgG are detected by the test in addition to a positive control band. Through the collection of this data, the COV-CLEAR study aims to address questions relating to  (a) the quantification of the robustness of the antibody response, and (b) the durability of this response. However, while affordable, we highlight that the LFA tests suffer from vastly varying levels of sensitivity --- registering a sensitivity as low as 70\% on asymptomatic individuals.
% \bgroup
% \def\arraystretch{1.5}
% \begin{table}[]
% \centering
% \begin{tabular}{lllllll}
% \hline
% \multicolumn{1}{|l|}{}           & \multicolumn{3}{c|}{Sensitivity}          & \multicolumn{3}{c|}{1- Specificity}   \\ \hline
% \multicolumn{1}{|l|}{}           & \multicolumn{3}{l|}{Mean [Low, High]}     & \multicolumn{3}{l|}{Mean [Low, High]} \\ \hline
% \multicolumn{1}{|l|}{Asymptomatic} & \multicolumn{3}{l|}{0.694 [0.519, 0.837]} & \multicolumn{3}{l|}{\multirow{4}{*}{0.005 [0.003, 0.008]}} \\ \cline{1-4}
% \multicolumn{1}{|l|}{2-10 days}  & \multicolumn{3}{l|}{0.811 [0.724, 0.881]} & \multicolumn{3}{l|}{}                 \\ \cline{1-4}
% \multicolumn{1}{|l|}{11-20 days} & \multicolumn{3}{l|}{0.935 [0.843, 0.982]} & \multicolumn{3}{l|}{}                 \\ \cline{1-4}
% \multicolumn{1}{|l|}{21+ days}   & \multicolumn{3}{l|}{0.98 [0.9, 1.0]}      & \multicolumn{3}{l|}{}                 \\ \hline
%                                  &               &              &            &             &             &          
% \end{tabular}
% \caption{Variations of sensitivity and specificity of immunoassay tests.}
% \label{tab:sens-spe}
% \end{table}
% \egroup

\xhdr{Clinical Data: Questionnaire} Additionally, participants are asked to answer a questionnaire, ideally before knowing the result of their LFA test. The form consists of questions related to $K=14$ exhibited symptoms (fever, cough, runny nose, etc.) and $M=2$ subject-specific risk factors (household size and proportion of members with a suspected or confirmed history of COVID-19). The empirical distributions of the symptoms and the risk factors in this cohort are provided in the supplementary materials. 
%Our goal is to leverage both subject-specific and global information on risk levels, as well as exhibited symptoms, to increase the precision of the diagnosis 
%\vspace{-0.3cm}

\section{Multimodal Data Integration: Statistical Model}\label{sec:multimodal}
\vspace{-0.3cm}

%We formalize our approach that performs multimodal data integration of medical data to improve diagnosis accuracy.
In this section, we describe our principled integration of noisy diagnostic test results, with additional clinical data such as symptom data and subject-specific risk factors. % -- in order to increase diagnosis accuracy. 
%Although we motivate our reasoning with LFA diagnostic tests for COVID-19, 
Our approach applies to general heterogeneous medical data where the outputs are binary. The latter could be either self-reported answers to questionnaires, clinician-reported physiological exams, outputs of a diagnostic based on an image, abnormalities of laboratory results, etc., making this a widespread and general setting. Thus, while we implement and showcase the method for the particular purpose of applying it to antibody testing, this could be relevant to any medical diagnostic with binary inputs.

Denote $D$ the true diagnosis of an individual (healthy/sick), $T$ the outcome of the noisy diagnostic test (positive/negative), $S$ the symptomaticity (symptomatic/asymptomatic), $X$ the symptoms exhibited and $Y$ the subject-specific risks factors. The underlying assumption is that given a true diagnosis $D$, the symptoms $X$ and the diagnostic test outcome $T$ are independent. In other words, the probability of the diagnostic test being a false negative $\mathbb{P}[T=0|D=1]$ is independent of the symptoms of a truly infected individual $\mathbb{P}[X|D=1]$. Similarly, given a diagnosis $D$, the test outcome $T$ and the exhibited symptoms $X$ are independent of the risk factors $Y$. We define:
\begin{itemize}[leftmargin=3em]
       \item $\p(D=1|Y) = \pi_\beta(Y)$ the probability of contracting the disease given risk factors $Y$,
       \item $\mathbb{P}[(T=1|D=1) = x$ the sensitivity of the diagnostic test,
       \item $\mathbb{P}[T=0|D=0] = 1-y$ the specificity of the diagnostic test, \textit{i.e.} $y=\mathbb{P}[T=1|D=0]$,
       \item $\mathbb{P}(S=1|D=0) = p_0$ the probability of being symptomatic when whilst not having been infected by that specific disease (the symptoms could be due to another illness for instance),
       \item $\mathbb{P}(S=1|D=1) = p_1$ the probability of having been symptomatic upon infection,
       \item $\mathbb{P}(X_k=1|S=1,D=0) = s_{0k}$ the probability of exhibiting symptom $k$ when not infected,
       \item $\mathbb{P}(X_k=1|S=1,D=1) = s_{1k}$ the probability of exhibiting symptom $k$ upon infection.
\end{itemize}
In the above, the diagnostic test may be any biological test (e.g. LFA antibody tests), or output of a medical imaging classification algorithm provided by a ML research team. We have here splitted our binary variables into two classes according to their variability: (a) the test $T$, for which we have provisional estimates for the sensitivity and specificity (given by the manufacturer) but which we are still uncertain, and (b) the symptoms $X$, which carry additional uncertainty in that neither of them is extremely specific to COVID-19 nor is their prevalence amongst COVID cohorts very well established; Values for $(x, y)$, $p_0, p_1, s_0, s_1$ or $\beta$ may be published, but challenged by complementary research or field experience, or may be completely unavailable.
We thus need to account for these inputs' variability and for the relative importance of different risk factors $\beta$. 
%We thus need to account for the uncertainty on the advertised sensitivity $x$ and specificity $y$ of the diagnostic test, as well as on the probabilities of symptomaticity $p_0, p_1$ and symptoms $s_0, s_1$ and on the relative importance of different risk factors $\beta$. 
We propose using a hierarchical Bayesian network (see Fig.~\ref{fig:model}) that is consequently deeper than the ones traditionally implemented in the CAD literature \cite{Alvarez2006EvaluationStenosis, Gevaert2006PredictingNetworks, Sakai2007AccuracyModel, Gonzalez-Lopez2016DevelopmentUveitis,Sangamuang2018AccuracyDecision}.
 
The uncertainty in the test sensitivity $x$ and specificity $1-y$ are expressed through Beta priors on $x$ and $y$ with parameters $(\alpha_{x}, \beta_x)$ and $(\alpha_y, \beta_y)$, see Fig.~\ref{fig:model}. Similarly, the uncertainty on the probabilities of symptomaticity and of the different symptoms are expressed via Beta priors with respective parameters $(\alpha_{p_0}, \beta_{p_0}), (\alpha_{p_1}, \beta_{p_1}), \{(\alpha_{s_{0k}}, \beta_{s_{0k}}), (\alpha_{s_{1k}}, \beta_{s_{1k}})\}_k$, see Fig.~\ref{fig:model}. When published estimates exist for $x, y, p_0, p_1, s_0$ and $s_1$ (e.g. as provided by the LFA test manufacturer), we match the moments of the Beta priors with those of the published distributions.
%In the COVID-19 example, we integrate the different values of sensitivity and specificity observed on the field for the LFA tests. 
If this information is unavailable, we use the non-informative Beta prior with parameters $(0.5, 0.5)$. In the COVID-19 example, we implement a non-informative prior for the probabilities of symptomaticity and symptoms, as the etiology of the disease remains unknown. These priors are then updated during learning as we aggregate the information from the observed and imputed variables.

Lastly, we model the probability $\pi_\beta(Y)$ of contracting the disease with a logistic regression on the $M$ risk factors $Y$. In the case of the COVID-19 dataset, these include county data, profession, size of household, etc. The coefficients $\beta=\{\beta_m\}_m$ weight the relative importance of each factor $Y_m$, $m=1...M$. We express our uncertainty on the possible impact of a given factor $m$ by introducing a Gaussian prior: $\beta \sim N(0, \sigma_\beta^2)$, see Fig.~\ref{fig:model}.

%The hierarchical Bayesian network is illustrated with plate representation in Figure~\ref{fig:model}. 

%In this model: (i) $\zeta = \Big(\alpha_x, \beta_x, \alpha_y, \beta_y, \alpha_{p_0}, \beta_{p_0}, \alpha_{p_1}, \beta_{p_1}, \{\alpha_{s_{0k}}, \beta_{s_{0k}}\}_k, \{ \alpha_{s_{1k}}, \beta_{s_{1k}}\}_k, \{\sigma_{\beta_m}\}_m\Big)$ are hyperparameters, considered fixed and known, (ii) $\theta = \Big(x, y, p_0, p_1, \{s_{0k}, s_{1k}\}_k, \{\beta_m\}_m \Big)$ are the parameters, (iii) $D$ is a hidden random variable, (iv) $Y, S, X, T$ are the observed variables. Further details on the model are available in the supplementary materials.

\begin{figure}[h!]
\centering
\def\svgwidth{0.8\columnwidth}
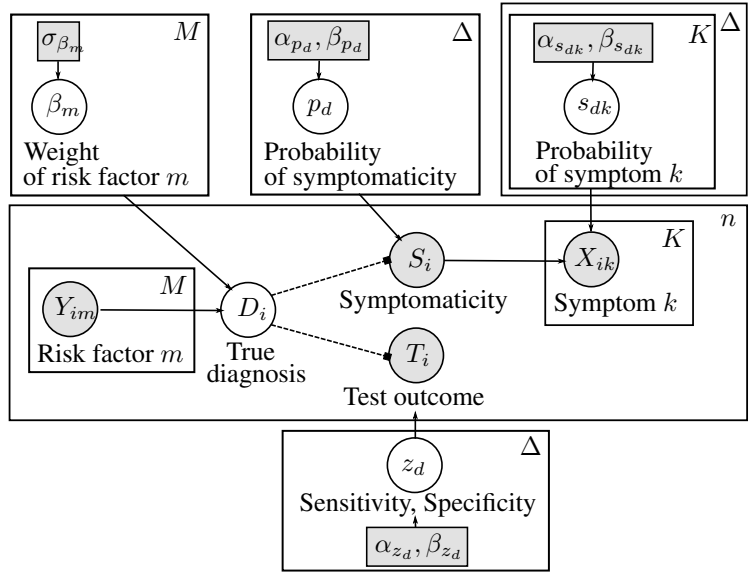
    \caption{Hierarchial bayesian network integrating the accuracy uncertainty on the data sources, to estimate the true diagnosis $D$. The index $k=1...K$ represents symptom $k$, while $m=1...M$ represents risk factor $m$. The shaded cells represent observed variables or known hyper-parameters. The dashed lines represent switches.}
    \label{fig:model}
\end{figure}
\vspace{-0.3cm}

% The likelihood of a given outcome $(D, X,T)$ under this model can thus be written as:

% \begin{equation}
%     \begin{split}
%     p(T,X, D)& = p_1(T)^{TD} (1-p_1(T))^{D(1-T)}  p_2(T)^{T(1-D)} (1-p_2(T))^{(1-D)(1-T)}\pi^D (1-\pi)^{1-D}  \\
%     &\times p(x|D=1)^D p(x|D=0)^(1-D) \times x^{\alpha_x-1} (1-x)^{\beta_x-1} y^{\alpha_y-1} (1-x)^{\beta_y-1}\\
%         \end{split}
% \end{equation}
\section{Inference in the Hierarchical Bayesian Network via Stochastic EM}
\vspace{-0.3cm}

\textit{At the subject level}, we seek to compute the posterior distribution of the true diagnosis $D_i$, informed by the integration of the observed variables $T_i, S_i, X_i, Y_i$. This posterior yields an estimated diagnosis, together with a credible interval that expresses the confidence associated with our prediction. In the case of COVID-19, this provides individual estimates of each citizen's immunity that may inform their social behavior as lockdown is eased.

\textit{At the global level}, we wish to learn: (i) the distributions of the sensitivity $x$ and specificity $y$ of the diagnostic test as observed in the field; (ii) the distributions of the probabilities $p_0, p_1$ of being symptomatic, and $s_0, s_1$ of exhibiting specific symptoms, within a population of interest; and (iii) the distribution of the impact $\beta$ of the risk factors for contracting the disease. In the context of the COVID-19 pandemic, such aggregated figures are pivotal to understand the dynamic of the disease and implement appropriate crisis policy.

Since the true diagnoses are hidden variables, the Expectation-Maximization algorithm \cite{Dempster1977MaximumAlgorithm} is an appealing method to perform inference in our Bayesian model; hence to achieve the aforementioned objectives. However, the EM algorithm requires the computation of an expectation over the posterior of the hidden variables, which may be intractable depending on the probability distributions defined in the model. To allow for flexibility in terms of model's design within our multimodal data integration framework, we offer to rely on the Stochastic EM algorithm (StEM) \cite{Celeux1988AAlgorithm}. StEM effectively estimates the conditional expectation in the EM using the ``Stochastic Imputation Principle", \textit{i.e.} approximating the expectation by sampling once from the underlying distribution. This method allows us to carry out inference with priors that are not necessarily the Beta distributions implemented in our experiments --- thus providing us with additional flexibility in modeling real data. Furthermore, StEM is more robust, being less dependent on the parameters' initialization than the EM -- a definite advantage given our very uncertain framework. Finally, StEM shows better asymptotic behavior: unlike the EM, StEM always leads to maximization of a complete data log-likelihood in the M-step \cite{Nielsen2000TheResults}. 
%The proofs for the propositions of this section are provided in the supplementary materials.

 % We choose the Stochastic EM algorithm (StEM)  over the Monte-Carlo EM (MCEM) algorithm \cite{Wei1990AAlgorithms}, as ...

%Since $D$ are hidden variables, we proceed with the Expectation-Maximization (EM) algorithm, specifically its stochastic version. We iterate the two steps of the stochastic EM algorithm, for a number of iterations $j=1...100$. 

%[RESULT OF SEM CONVERGENCE HERE.]

% in contrast our approach uses a Stochastic Expectation Maximization algorithm, that allow for additional flexibility as to which exponential family we use. Moreover, since we are in an even higher-uncertainty level than most studies,  using the Stochastic version of EM helps, since this has been shown to be more robust against bad initialization.
\vspace{-0.3cm}

% Question: Shouldn't we iterate until convergence, instead of setting a total number of iterations?

\subsection{Stochastic E-step at iteration $(j+1)$}
\vspace{-0.1cm}

Starting iteration $j+1$, the current estimates of the model's parameters are $\theta^{(j)}$. The stochastic E-step computes the posterior distribution $\p(D_i | T_i, S_i, X_i, Y_i, \theta^{(j)})$ of the hidden variable $D_i$. We then sample a diagnosis $\widehat{D_i}$ from this posterior.

\begin{proposition}\label{prop:d}
The odds of the posterior of the hidden variable $D_i$ at iteration $(j+1)$ writes:
\begin{equation}
   \begin{split}
    &\frac{
        \p(D_i=1|T_i, S_i, X_i, Y_i, \theta^{(j)})
        }{
        \p(D_i=0|T_i, S_i, X_i, Y_i, \theta^{(j)})}\\
    &= \frac{
    {x^{(j)}}^{T_i} (1-x^{(j)})^{(1-T_i)}
    \times {s_1^{(j)}}^{S_iX_i} (1-s_1^{(j)})^{S_i(1-X_i)} 
    \times \pi(Y_i, \beta^{(j)})
    \times p_1^{S_i}(1-p_1)^{(1-S_i)}
    }{
    {y^{(j)}}^{T_i}(1-y^{(j)})^{(1-T_i)}\times {s_0^{(j)}}^{S_iX_i}(1-s_0^{(j)})^{S_i(1-X_i)} \times (1-\pi(Y_i, \beta^{(j)}))
        \times p_0^{S_i}(1-p_0)^{(1-S_i)}}.
    \end{split}
\end{equation}
\end{proposition}

\subsection{M-step at iteration $(j+1)$}
\vspace{-0.1cm}

The following proposition shows the updates in the model's parameters at iteration $(j+1)$, performed in the M-step.

\begin{proposition}\label{prop:xy}
The parameters updates write:
\begin{equation}
    \begin{gathered}
       x^{(j+1)}  
       = \frac{
       \sum_{i=1}^nT_i\widehat{D_i}+\alpha_x - 1}{
       \sum_{i=1}^n\widehat{D_i}+(\alpha_x+\beta_x) - 2 }, \quad
       y^{(j+1)}
       = \frac{
       \sum_{i=1}^nT_i(1-\widehat{D_i})+\alpha_y - 1}{
       \sum_{i=1}^n (1-\widehat{D_i})+(\alpha_y+\beta_y) - 2}\\
       p_0^{(j+1)} 
       = \frac{
       \sum_{i=1}^n (1-\widehat{D_i}) S_i + \alpha_{p_0} - 1}{
       \sum_{i=1}^n (1 - \widehat{D_i})+(\alpha_{p_0}+\beta_{p_0}) - 2 }, \quad
       p_1^{(j+1)} 
        = \frac{
       \sum_{i=1}^n \widehat{D_i}S_i + \alpha_{p_1} - 1}{
       \sum_{i=1}^n \widehat{D_i}+(\alpha_{p_1}+\beta_{p_1}) - 2}\\
       s_0^{(j+1)} 
       = \frac{
       \sum_{i=1, \text{s.t. } S_i=1}^n (1-\widehat{D_i}) X_i + \alpha_{s_0} - 1}{
       \sum_{i=1, \text{s.t. } S_i=1}^n (1 - \widehat{D_i})+(\alpha_{s_0}+\beta_{s_0}) - 2 }, \quad
       s_1^{(j+1)} 
        = \frac{
       \sum_{i=1, \text{s.t. } S_i=1}^n \widehat{D_i}X_i + \alpha_{s_1} - 1}{
       \sum_{i=1, \text{s.t. } S_i=1}^n \widehat{D_i}+(\alpha_{s_1}+\beta_{s_1}) - 2}\\
       \beta^{(j+1)}
        = \text{argmin}_\beta \sum_{i=1}^n \frac{||\widehat{D_i} - g(Y_i\beta)||^2}{2\sigma^2} + \frac{||\beta||^2}{2\sigma_\beta^2},
    \end{gathered}
\end{equation}
where the minimization on $\beta$ is performed through Newton-Raphson descent.
\end{proposition}

\xhdr{Algorithm Complexity} Our algorithm only relies on simple sequential updates of the distribution parameters, as highlighted in Prop.~\ref{prop:d} and \ref{prop:xy}. Denoting by $L$ the number of such parameters, and $N$ the number of samples:
\vspace{-0.4cm}

\begin{itemize}[leftmargin=2em]
\setlength\itemsep{0em}
    \item The updates for $X$ and $T$ only involve matrix multiplications of the form $\widehat{D}^TX$ , while those for $\beta$ involve matrix multiplications of size $O(M^2N)$ --- yielding a complexity of $O(LN)$.
    \item Prior updates rely solely on element-wise operations on the log-odds, with $O(N)$ complexity.
\end{itemize} 
Denoting as $B$ the maximum number of steps, the complexity is $O(BLN)$, and thus linear in $N$. Fig~\ref{fig:bench_perf}(A) provides an example of the evolution of the number of steps as the number of samples increases.
%We note that the EM algorithm typically converges in a few steps. 

\vspace{-0.2cm}

\section{Validation on Synthetic Data}\label{sec:simu}
\vspace{-0.3cm}

Since our approach is unsupervised, we begin by validating it on synthetic datasets where the ground truth is known and controlled --- thus allowing us to characterize the performance of the algorithm, and to showcase the strength of combining multiple noisy modalities. 
%In particular, we aim to assess the accuracy of the recovery of (a) the ground truth diagnosis $D$ and (b) the different parameters of the models as a function of the number of samples, specificity, and sensitivity.

 We assume the same generative model as in Fig. \ref{fig:model} and generate synthetic data for various pairs of values of sensitivity and specificity ranging from 60\% to 99\% . In each case, we simulate $N=100$ tuples  of variables ($(Y_i, X_i, T_i)_{i=1}^{n}$, for $n \in \{100, 200, 500, 1000\}$ participants. We also vary the noise level $\sigma \in \{0.1, 0.5, 1.0\}$ for the prior on $D$ in  Fig. \ref{fig:model}. To mimic our COVID data, we simulate $K=14$ symptoms and $M=2$ risk factors, for which we randomly select the parameters\footnote{All of these parameters are provided with the code in the supplementary materials.}. 
 %The results of our Stochastic EM algorithm (StEM) are presented as follows.

\begin{figure}
\centering
    \includegraphics[width=\textwidth]{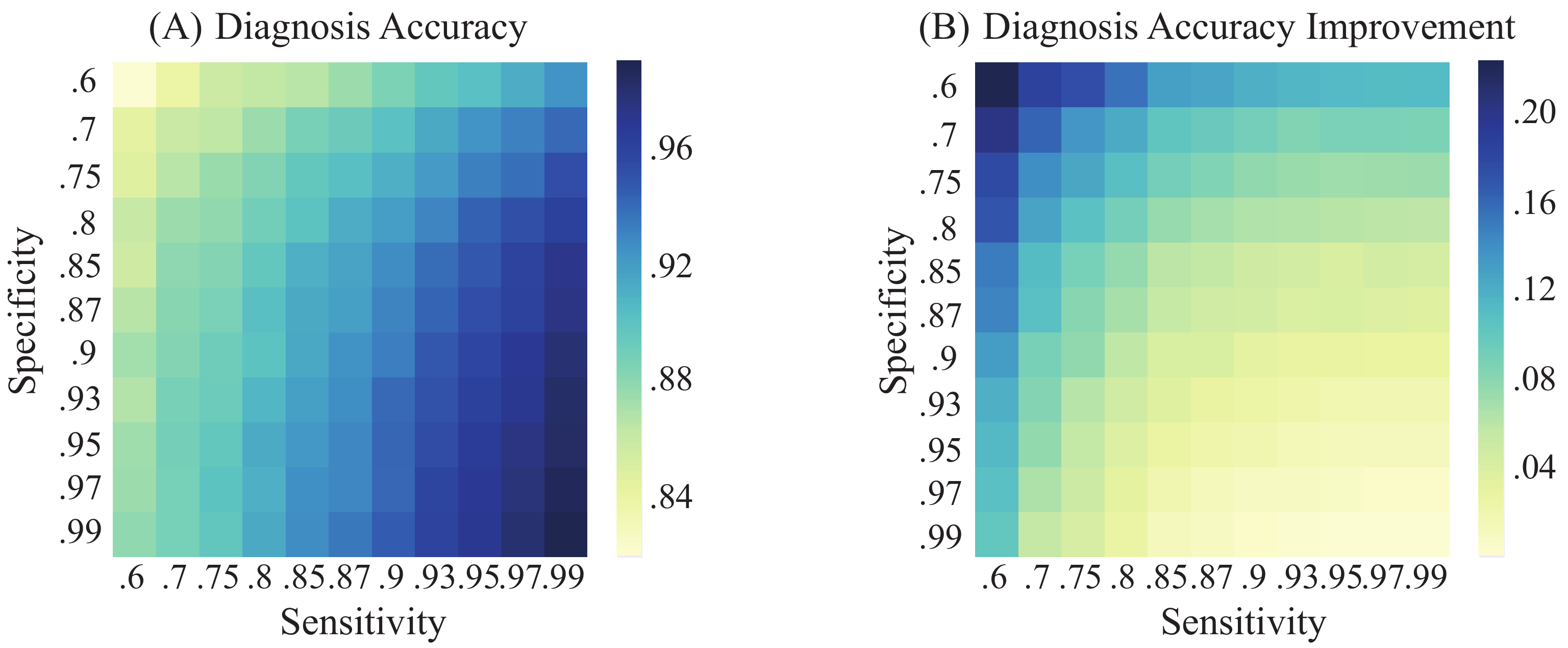}
    \caption{Performance of the SEM algorithm for n=300 samples, $\sigma=0.5$, and varying levels of specificity and sensitivity. \textbf{(A)} Raw accuracy of the labels imputed via StEM. \textbf{(B)} Difference in accuracy between the StEM imputed diagnostics and the observed test variable $T$.  }
    \label{fig:heatmap_acc}
\end{figure}

\xhdr{Improving upon the sole test $T$} Fig.~\ref{fig:heatmap_acc} (A) and (B) show the Stochastic EM's raw accuracy and accuracy improvement over the sole test result $T$. Note that this difference is always positive, highlighting that our method only improves upon single diagnostic inputs --- even when one input is more reliable than any of the others.
%, and is by no means hindered by the combination of all heterogeneous data sources. 
As expected, the most substantial improvements upon $T$ are observed when the test specificity and/or sensitivity are low. For instance, for sensitivity and specificity of 70\%, our method provides an  $86.6 \pm (2.5)$\% accuracy for the diagnosis -- or equivalently, a 16\% gain over $T$.
%: the test's accuracy is $69.8 \pm 3.2$ while the algorithm is at $86.6 \pm 2.5 $.
We further quantify the algorithm's performance in Table 1 of the supplementary materials, for values of the specificity close to those observed on the field.

% \begin{table}[ht]
% \caption{Gain in accuracy (mean and standard deviation) when using StEM over the sole test}
% \begin{center}
% \begin{tabular}{|c|c|cccc|}
%     \hline
%     \multicolumn{2}{|c|}{Sensitivity}& \multicolumn{4}{|c|}{Specificity}\\\hline \hline
%     & &  70 & 80 & 93 & 99\\ \hline 
%     Asymptomatic & 69.4 & 16.2 $\pm 3.9$ & 12.8 $\pm 4.0$ & 8.42 $\pm 3.2$&  5.64 $\pm 3.2$\\ \hline
%     2-10 days & 81.1 & 12.3 $\pm 3.2$ & 9.27 $\pm 3.1$& 4.82$\pm 2.9$&97.0 $\pm 2.3$\\ \hline
%     11-20 days & 93.5 & 8.5 $\pm 2.6$ & 6.4 $\pm 2.5$& 2.3 $\pm 2.1$&0.03 $\pm 1.4$\\ \hline
%     21+ days & 99.0  & 8.7 $\pm 2.7$ & 5.7 $\pm 2.0$& 2.0 $\pm 1.4$& 0.01 $\pm 0.0$\\ \hline
% \end{tabular}
% \end{center}
% \label{tab:multicol}
% \end{table}

%Tables~\ref{tab:res-asympt}-\ref{tab:res-21-plus} show the mean diagnosis accuracy obtained in the different cases, as well as the improvement with respect to the test mean accuracy in parenthesis. The distributions of diagnosis and test accuracies for the $N=100$ simulations are provided in the supplementary materials.

\xhdr{Benchmarking against other models} We now benchmark our algorithm against other approaches:
\vspace{-0.3cm}

\begin{itemize}[leftmargin=2em]
\setlength\itemsep{0em}
\setlength\topsep{0em}
    \item \textit{Vanilla Classifier:} using  both the context $Y$ and symptoms $X$ as inputs, we fit a logistic regression to the test labels $T$. We choose the regularization parameter using 10-fold cross-validation, and compute confidence intervals for the log-probability of $(D|X,T)$ by bootstrapping.
    \item \textit{Data-Agnostic EM}: we implement a deterministic version of EM, providing uninformative priors for the parameter (thereby reflecting our absence of knowledge of the truth), which are not updated --- i.e, an "uninformed'' equivalent of the belief networks found in the literature.
        \item \textit{Data-Informed EM}: similar to the Data-Agnostic EM, but choosing the priors (which then remain fixed) based on the empirical data.
\end{itemize}
\vspace{-0.3cm}

The results are shown in Fig.~\ref{fig:bench_perf}(B), and further completed in the supplementary materials. We highlight the superiority of our deeper and more adaptive hierarchical model, yielding improvements  (for a reasonable tuple of specificity and sensitivity of 80\%) of up to 4\% over the Data-Informed EM, 8\% over the Data-Agnostic one, and 9\% over the Vanilla classifier.
%In particular, we observe a X\% increase in StME over the EMs.

\begin{figure}
\centering
    \includegraphics[width=\textwidth]{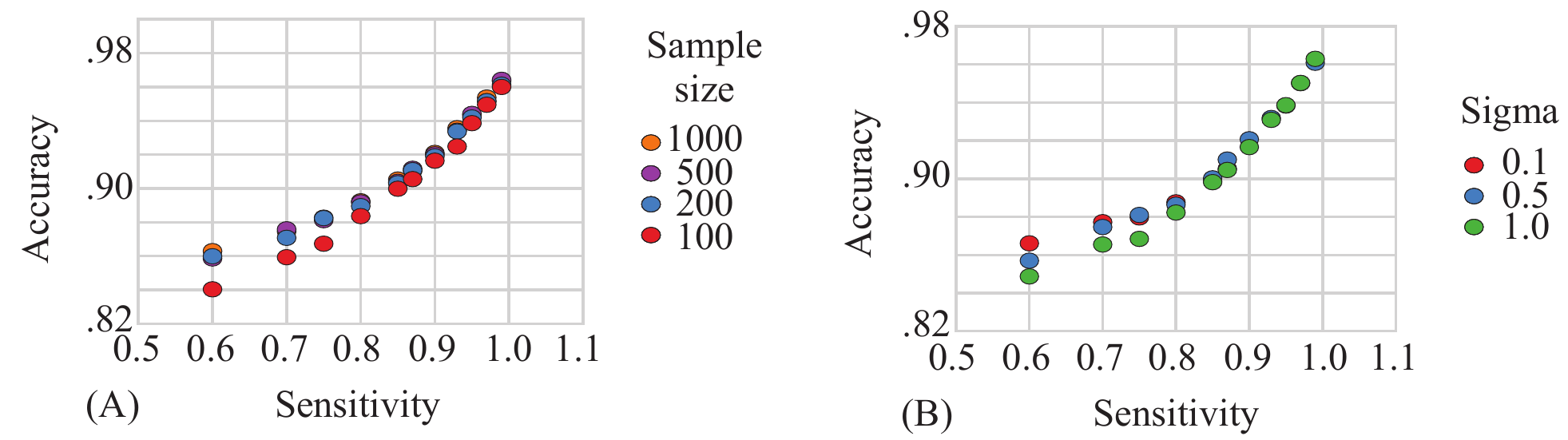}
    \caption{Performance of the Stochastic EM for a fixed specificity of 80\%,  (A) as the number of samples increases (and $\sigma=0.5$) and (B)  as the variance increases  (and $n=300$).}
    \label{fig:varying_inputs}
\end{figure}

\xhdr{Assessing the robustness of the Stochastic EM} For a fixed specificity 80\%, Figure \ref{fig:varying_inputs} shows the accuracy of StEM for different values of $\sigma$ (A) or as the number of samples increases (B). These axes of variation seem to yield little impact on the model's performance, providing evidence that our algorithm is fairly robust, especially with respect to low-sample size. %In other words, the performance of StEM is not severely impacted in low-sample regimes ($n\approx 100$) compared to high-sample ones.
%The results show a significant improvement of the test accuracy, demonstrating that using questionnaire data indeed enhances the immunity prediction from the immunoassay tests.

\begin{figure}
\centering
    \includegraphics[width=\textwidth]{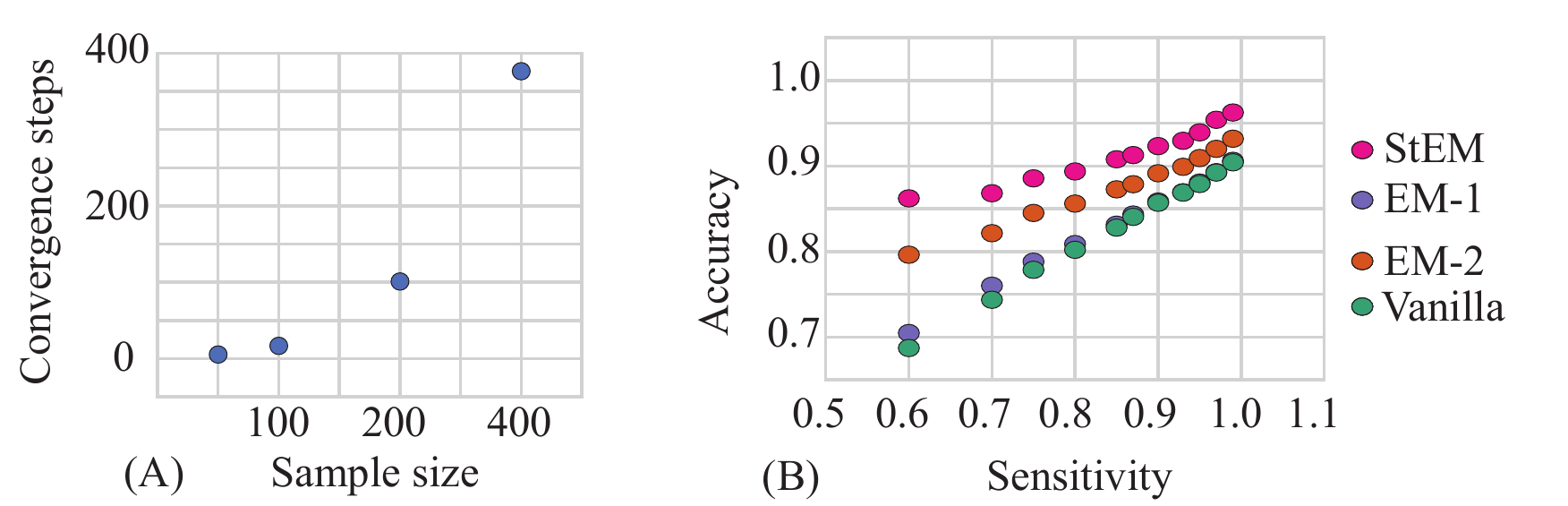}
    \caption{(A) Number of EM steps until convergence; and (B) Accuracy comparison against benchmarks: Stochastic EM (StEM), EM with data-informed priors on models' parameters, not updated during learning (EM-2), EM with uninformative priors on models' parameters, not updated during learning (EM-3), Vanilla logistic regression (Log. Reg.).}
    \label{fig:bench_perf}
\end{figure}

\xhdr{Convergence} Finally, we assess the convergence properties of our algorithm. We examine the distribution of the relative difference between recovered coefficients and ground truth (expressed as a percentage of the ground-truth value). The plots, provided in the supplementary materials, show deviations that are within a few percentages of the true value of the coefficients -- thus highlighting the ability of the model to converge to the ground-truth parameters, and making it relevant from a medical perspective to characterize the disease's etiology. Illustrations of the behavior of the number of required EM steps are also provided in the supplementary materials.
%The fact that these parameters converge to the ground truth ones renders them interpretable from the medical perspective.

% \begin{figure}
% \centering
%     \includegraphics[width=0.49\textwidth]{FIG_SIM/sensitivity_MSE_beta.pdf}
%     \includegraphics[width=0.49\textwidth]{FIG_SIM/accuracy_recovery.pdf}
%     \caption{(a) MSE for the beta vs (b) MS for the parameters of the mdoels }
%     \label{fig:heatmap_acc}
% \end{figure}

\vspace{-0.3cm}

\section{Results on Real Data}\label{sec:real}
\vspace{-0.3cm}

We turn to the real dataset of $n=117$ participants described in Section~\ref{sec:data}. The purpose of this section is to show how our algorithm can be practically applied to process heterogeneous data types and inform participants and researchers alike. At the individual level, we provide each user with (a) the confirmation of the diagnostic, and (b) confidence intervals associated with the uncertainty to shed more light on the uncertainty associated with the diagnostic. At the global level, we provide policymakers with (c) aggregated analysis of the herd immunity, and associated measures of uncertainty. 

\begin{figure}
    \centering
    \includegraphics[width=\textwidth]{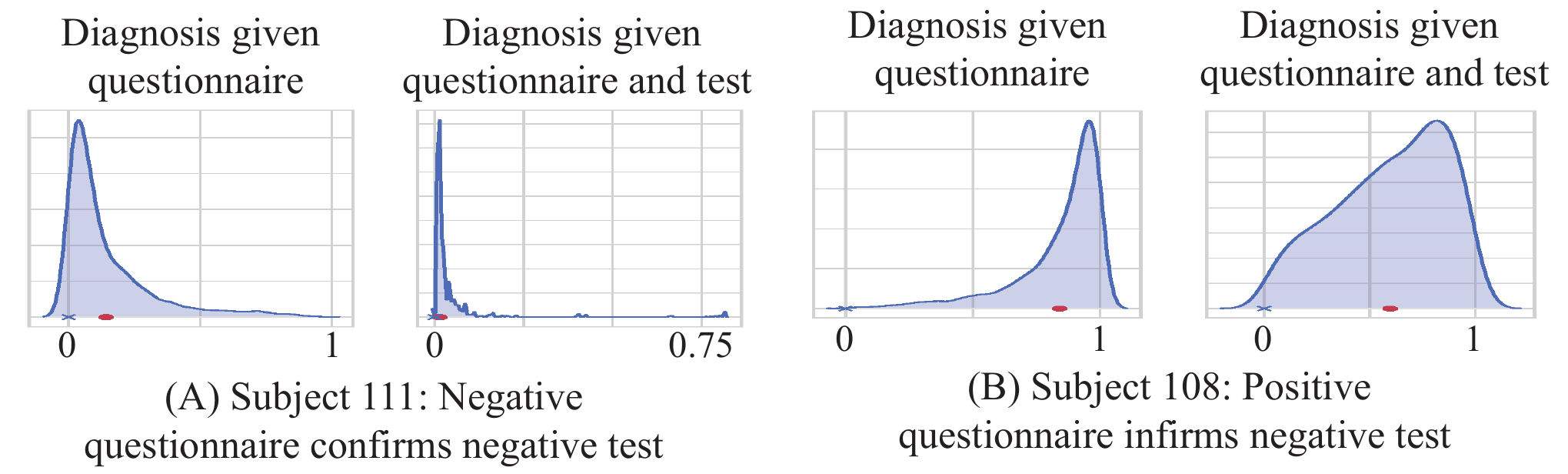}
    \caption{Posterior diagnosis distributions on two selected subjects. In each panel (A-B): the left plot represents the posterior of the diagnosis, given the symptoms and risk factors data reported in the questionnaire, while the right plot is the posterior of the diagnosis, given the symptoms, risk factors, and reported result of the diagnostic test. The red dot represents the expectation.}
    \label{fig:real}
\end{figure}

\xhdr{At the subject-level} We present two examples, where our algorithm either confirms or infirms the result of the test -- thereby allowing for the potential flagging of false negatives. The first example (Fig.~\ref{fig:real} A) is a user that registers a negative test while being asymptomatic and with a limited number of risk factors. In this case, we expect our model to confirm the result of the test, and provide a narrower confidence interval as per the probability of immunity -- as confirmed by Fig.~\ref{fig:real} (A). The second example showcases an instance where the questionnaire and the test disagree. Subject 108 is a user that registers a negative test, while exhibiting a wide number of known COVID symptoms (dry cough, shortness of breath, fever), but took the test less than 10 days after his illness. While the posterior does not reclassify the subjects' diagnosis, the confidence interval associated with the prediction of immunity reflects the uncertainty associated with this case, and flags it as a potential false negative.

%h specificity of the test (99\%), the probability of encountering cases where we would want to infirm positive tests is extremely small.
%highlighted by the extreme uncertainty of his posterior diagnosis.
%This could be of help to the subject, encouraging them to try a test a second time, and/or putting the appropriate weight on this particular test outcome when computing prevalence rates among a given population.

%\vspace{-0.3cm}

\begin{figure}
\centering
    \includegraphics[width=\textwidth]{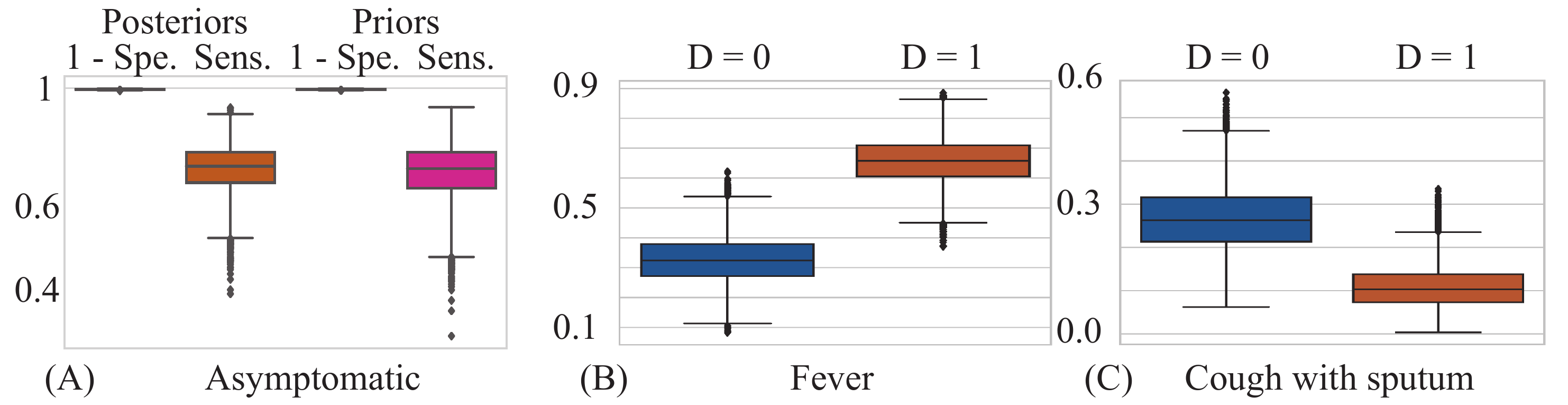}
    \caption{(A) Posteriors of sensitivity and specificity compared to their priors, for asymptomatic subjects. (B-C): Posteriors of the probability of exhibiting specific symptoms, for symptomatic subjects with estimated negative ($D=0$) or positive ($D=1$) diagnosis.}
    \label{fig:global}
\end{figure}

\xhdr{At the global level} At the global level, this framework allows to perform principled inference for the disease and the population of interest.
%, in our case a cohort of $n=117$ healthcare workers in the UK.
For this  cohort of $n=117$ healthcare workers in the UK, at-home tests predict $35.8\%$ of immunity. 
%However, taking into account the uncertainties associated with these tests, our method predicts that, in fact, $X\%$ of this population is immune, with confidence interval $[X, X]$. 
The posterior distributions of the model's parameters shed light on the actual accuracy of the LFA tests on our population. Fig.~\ref{fig:global} (A) compares the posteriors of the sensitivity and specificity to the priors built from values reported by manufacturers, for asymptomatic subjects. Furthermore, our results provide information regarding the most prevalent symptoms for COVID-19. For instance, among symptomatic participants, the probability of exhibiting fever is higher for infected subjects (see Fig.~\ref{fig:global} (B)) while cough with sputum does not seem to be associated with COVID-19, being probably the sign of another infection (see Fig.~\ref{fig:global} (C)). Similar plots on additional symptoms are provided in the supplementary materials.

% TODO: group by D=0, D=1 and not T=0, T=1 -> compare
\vspace{-0.2cm}

\vspace{-0.3cm}
\section{Conclusion}
\vspace{-0.3cm}

This applications paper provides a statistical framework for multimodal integration of noisy diagnostic test results, self-reported symptoms, and demographic variables. Compared to previous approaches, our work is more amenable to the handling of the different inputs' uncertainty. While we have focused here on binary symptoms, this adaptive Bayesian framework, together with the flexibility provided by the Stochastic EM algorithm, pave the way for the integration of other continuous-valued inputs (index of the severity of the disease, pain, etc.) as well as interactions -- an extension that we are currently in the process of implementing. The robustness of the Stochastic EM is crucial in ensuring the reliability of the parameters given the medical stakes. Finally, we note that the generative model allows for principled handling of missing data, thus allowing us to make diagnoses even in the presence of incomplete data.

\appendix

\section{COVID-19 Datasets}

\begin{figure}[h!]
\centering
\includegraphics[width=\textwidth]{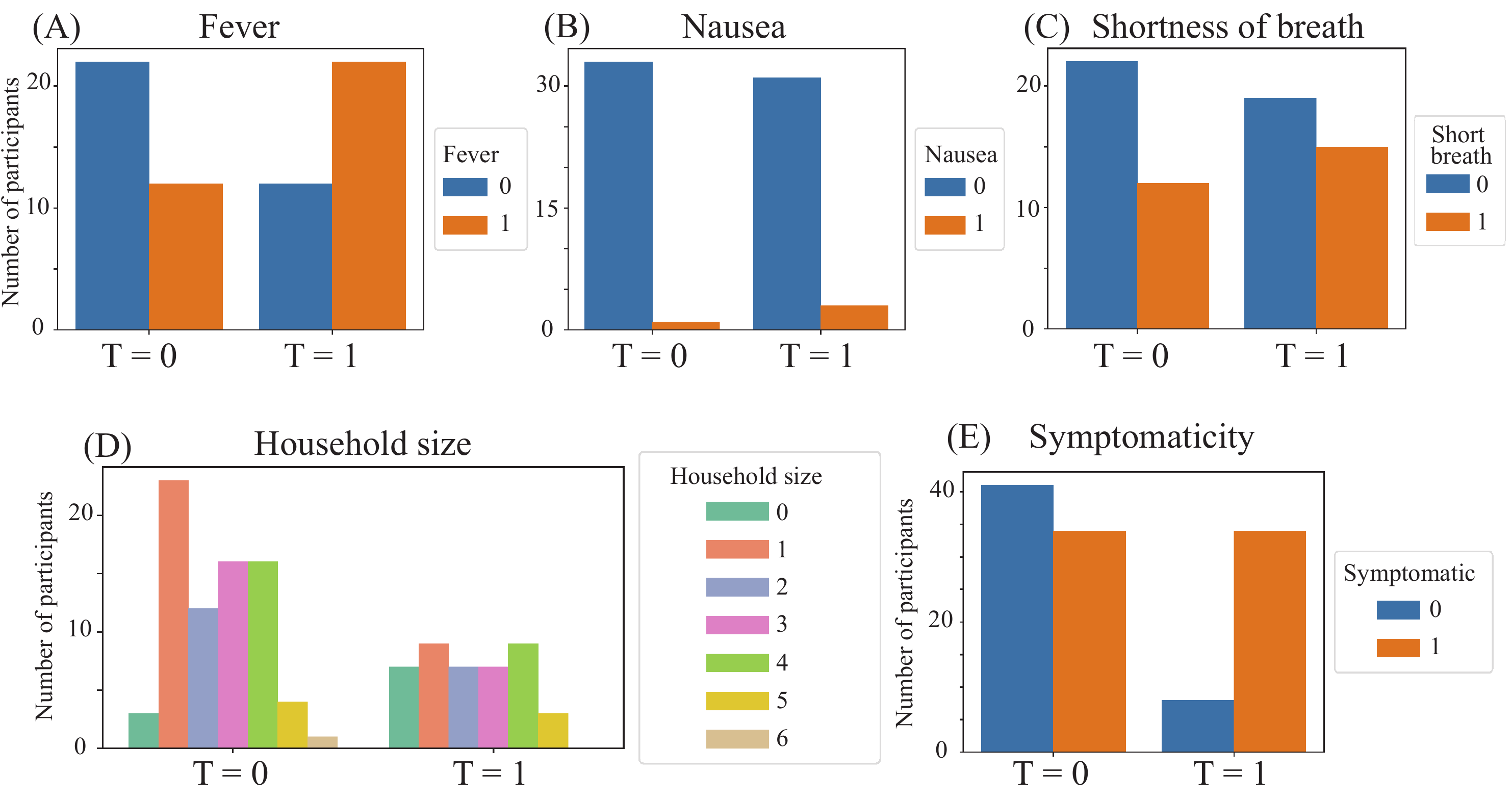}
    \caption{(A-C) Distributions of selected symptoms in the COV-CLEAR dataset. (D) Distribution of household size, a COVID-19 risk factor, in the COV-CLEAR dataset. (E) Distribution of symptomatic subjects in the COV-CLEAR dataset. In each plot, the distributions are grouped with respect to the result of the LFA test: $T=0$ for a negative test, and $T=1$ for a positive test.}
    \label{fig:symptoms}
\end{figure}

\begin{figure}[h!]
\centering
\includegraphics[height=2.5cm]{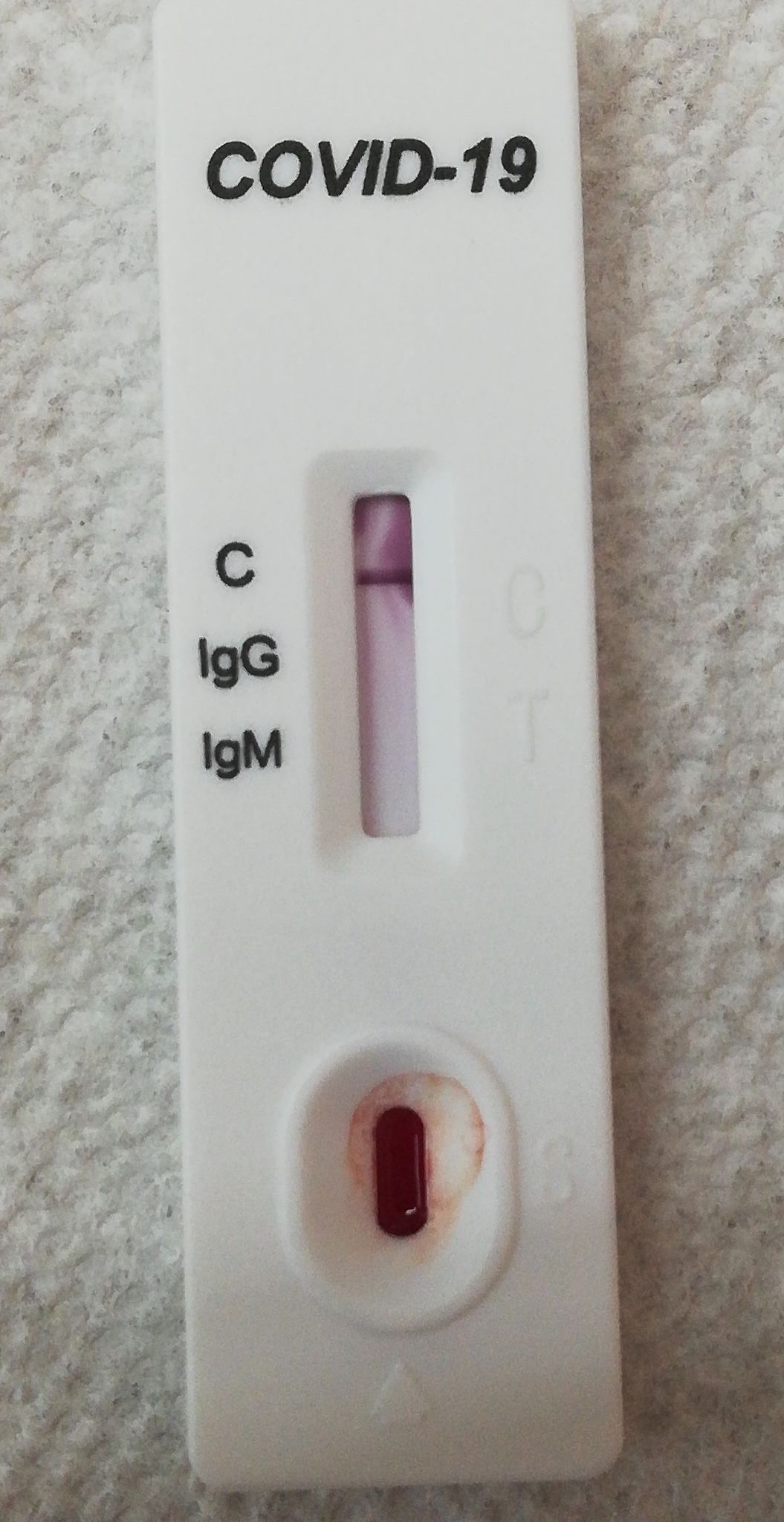}
\includegraphics[height=2.5cm]{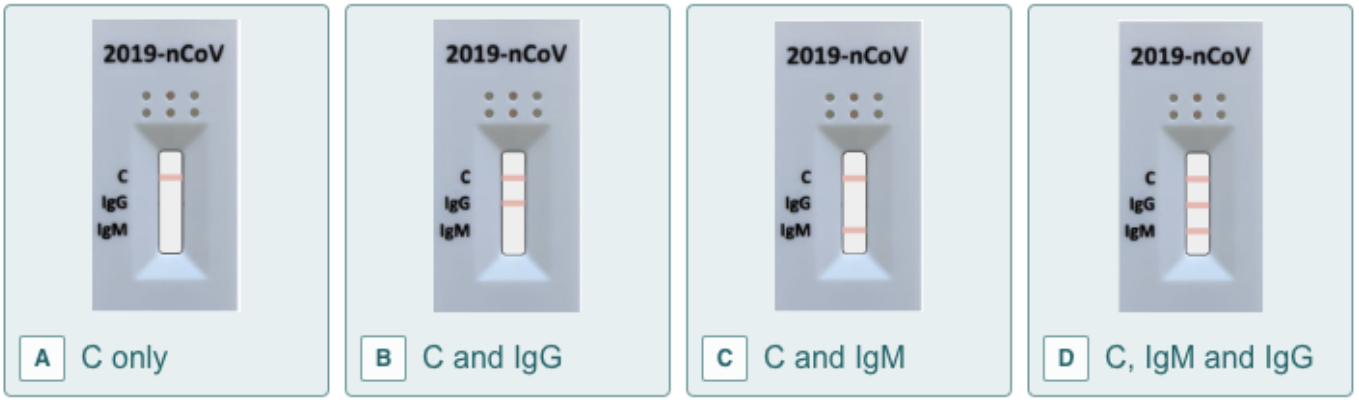}
    \caption{Left: Example of test result. Right: Illustrations of a subset of possible test results.}
    \label{fig:test}
\end{figure}

This section provides additional details on the COV-CLEAR dataset \footnote{\url{www.cov-clear.com}}. Figure~\ref{fig:symptoms} shows the sampling distributions of selected symptoms and risk factors in the population of interest -- grouped by the result of the lateral flow immunoassay (LFA) test. Figure~\ref{fig:test} illustrates the LFA test results as observed by a given subject. Participants self-report a positive result if IgM and/or IgG are detected by the test. The marker ``C" stands for ``Control". 

\section{Stochastic Expectation-Maximization}

This section presents the derivation of the Stochastic Expectation-Maximization (StEM) algorithm, specifically the proofs of Propositions 4.1 and 4.2  of the main paper. For the convenience of the reader, we recall the Bayesian model of interest. 

\subsection{Bayesian generative model}

Denote $D$ the true diagnosis of an individual (healthy/sick), $T$ the outcome of the noisy diagnostic test (positive/negative), $S$ the symptomaticity (symptomatic/asymptomatic), $X$ the symptoms exhibited and $Y$ the subject-specific risks factors. The underlying assumption is that given a true diagnosis $D$, the symptoms $X$ and the diagnostic test outcome $T$ are independent. In other words, the probability of the diagnostic test being a false negative $\mathbb{P}[T=0|D=1]$ is independent of the symptoms of a truly infected individual $\mathbb{P}[X|D=1]$. Similarly, given a diagnosis $D$, the test outcome $T$ and the exhibited symptoms $X$ are independent of the risk factors $Y$. We define:
\begin{itemize}[leftmargin=3em]
       \item $\p(D=1|Y) = \pi_\beta(Y)$ the probability of contracting the disease given risk factors $Y$,
       \item $\mathbb{P}[(T=1|D=1) = x$ the sensitivity of the diagnostic test,
       \item $\mathbb{P}[T=0|D=0] = 1 - y$ the specificity of the diagnostic test, \textit{i.e.} $y=\mathbb{P}[T=1|D=0]$,
       \item $\mathbb{P}(S=1|D=0) = p_0$ the probability of being symptomatic when whilst not having been infected by that specific disease (the symptoms could be due to another illness for instance),
       \item $\mathbb{P}(S=1|D=1) = p_1$ the probability of having been symptomatic upon infection,
       \item $\mathbb{P}(X_k=1|S=1,D=0) = s_{0k}$ the probability of exhibiting symptom $k$ when not infected,
       \item $\mathbb{P}(X_k=1|S=1,D=1) = s_{1k}$ the probability of exhibiting symptom $k$ upon infection.
\end{itemize}
 
The uncertainty in the test sensitivity and specificity is expressed by putting a prior on $x$ and $y$, which will be updated during training:
\begin{equation}
    \begin{split}
        \mathbb{P}[T=1|D=1] = x & \sim \Beta(\alpha_x, \beta_x),\\
       \mathbb{P}[T=1|D=0] = y & \sim \Beta(\alpha_y, \beta_y).
    % \log(\frac{\pi}{1-\pi})  &\sim N( \alpha + Y\beta, \sigma^2)  \text{\hspace{1cm} (Standard GLM)} \\
       %\epsilon &\sim \frac{1}{\Gamma(1)}
    \end{split}
\end{equation}

The uncertainty on the symptomaticity, as well as on the appearance of specific symptoms for infected and healthy individuals is expressed with a prior on $s_0$ and $s_1$, and updating it when we aggregate more information:
\begin{equation}
    \begin{split}
        \mathbb{P}[S=1|D=0] = p_0 & \sim \Beta(\alpha_{p_0}, \beta_{p_0}),\\
       \mathbb{P}[S=1|D=1] = p_1 & \sim \Beta(\alpha_{p_1}, \beta_{p_1}).\\
       \mathbb{P}[X=1|D=0] = s_0 & \sim \Beta(\alpha_{s_0}, \beta_{s_0}),\\
       \mathbb{P}[X=1|D=1] = s_1 & \sim \Beta(\alpha_{s_1}, \beta_{s_1}).\\
    \end{split}
\end{equation}

Lastly, we model the probability for an individual to contract the disease depending on their risk factors, by modeling the logodds:
\begin{equation}
    \log\left(\frac{\pi(Y)}{1-\pi(Y)}\right) = Y\beta + \epsilon \quad \text{where: } \epsilon \sim N(0, \sigma^2),
\end{equation}
where the parameters $\beta$ weight the importance of the different components of the risk factors $Y$ (county data, profession, size of household) in contracting the disease. We express our uncertainty on whether a given factor has an impact by putting a prior on $\beta$:
\begin{equation}
     \beta \sim N(0, \sigma_\beta^2).
\end{equation}
%Then, $\epsilon$ represents other factors that lead to the contraction of the disease that may not be taken into account here, as well as the uncertainty of the prediction model.

In this model:
\begin{itemize}
    \item $\zeta = \Big(\alpha_x, \beta_x, \alpha_y, \beta_y, \alpha_{p_0}, \beta_{p_0}, \alpha_{p_1}, \beta_{p_1}, \{\alpha_{s_{0k}}, \beta_{s_{0k}}\}_k, \{ \alpha_{s_{1k}}, \beta_{s_{1k}}\}_k, \{\sigma_{\beta_m}\}_m\Big)$ are hyper-parameters, considered fixed and known,
    \item $\theta = \Big(x, y, p_0, p_1, \{s_{0k}, s_{1k}\}_k, \{\beta_m\}_m \Big)$ are the parameters,
    \item $D$ is a hidden random variable,
    \item $Y, S, X, T$ are the observed variables. 
\end{itemize}

For simplicity of notations, we write $O = (S, X, T)$ some of the observed variables. The Bayesian model is represented in plate notations in Figure~\ref{fig:model}.

\begin{figure}[h!]
\centering
\def\svgwidth{0.8\columnwidth}
\input{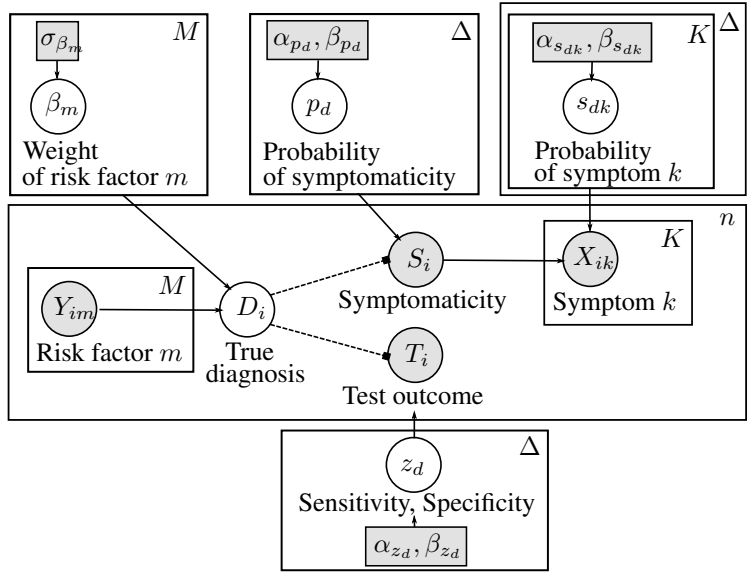}
    \caption{Hierarchial bayesian network integrating the accuracy uncertainty on the data sources, to estimate the true diagnosis $D$. The index $k=1...K$ represents symptom $k$, while $m=1...M$ represents risk factor $m$. The shaded cells represent observed variables or known hyper-parameters. The dashed lines represent switches.}
    \label{fig:model}
\end{figure}

% The likelihood of a given outcome $(D, X,T)$ under this model can thus be written as:

% \begin{equation}
%     \begin{split}
%     p(T,X, D)& = p_1(T)^{TD} (1-p_1(T))^{D(1-T)}  p_2(T)^{T(1-D)} (1-p_2(T))^{(1-D)(1-T)}\pi^D (1-\pi)^{1-D}  \\
%     &\times p(x|D=1)^D p(x|D=0)^(1-D) \times x^{\alpha_x-1} (1-x)^{\beta_x-1} y^{\alpha_y-1} (1-x)^{\beta_y-1}\\
%         \end{split}
% \end{equation}

Our objective is two-fold. \textit{At the patient level}, we compute the posterior distribution of the true diagnosis $D_i$, informed by the integration of the observed variables $O_i$. This distribution gives us an estimate of the true diagnosis, through Maximum a Posteriori, as well as a credible interval that expresses our confidence in this diagnosis. \textit{At the global level}, we learn the posterior distributions of the parameters $x, y$, i.e. the sensitivity and specificity of the test - to be compared to the values given by the providers. We also learn the posterior distributions of the parameters $s_0, s_1$, i.e. the probability of symptoms with or without the disease - to be compared to the values given by the medical specialists and the CDC. We also learn the posterior distribution of the parameter $\beta$, which weights the impact of each risk factor for contracting the disease.

To fulfil this objective, we perform inference in the Bayesian model described in Figure~\ref{fig:model} and in the main paper. Since $D$ are hidden variables, we proceed with the Expectation-Maximization (EM) algorithm, specifically its stochastic version.

\subsection{Stochastic Expectation-Maximization}

We seek the Maximum a Posteriori (MAP) of the parameters $\theta$. Given the parameters' estimates, we can compute the posterior distributions of the diagnosiss $D_i$'s. The stochastic EM algorithm allows computing jointly the posterior distribution of the parameters and the hidden variables $D_i$'s, through an iterative procedure described below.

We want to maximize the posterior distribution of the parameters $\theta$:
\begin{equation}
\begin{split}
        \p(\theta |O_1, ... O_n, Y_1, ..., Y_n) 
             = \frac{\p(O_1, ..., O_n|\theta, Y_1, ..., Y_n) \times \p(\theta)
                }{
                \p(O_1, ..., O_n|Y_1, ..., Y_n)} 
             \propto \p(\theta) \times \Pi_{i=1}^n \p(O_i|\theta, Y_i) 
\end{split}
\end{equation}
which translates into maximizing the expression:
\begin{equation}
    \begin{split}
  \ell 
  & = \sum_{i=1}^n \log \p(O_i|\theta, Y_i) 
      + \log \p(\theta) \\
  & = \sum_{i=1}^n \log \sum_{d_i=0,1} \p(O_i, D_i=d_i|\theta, Y_i)
      + \log \p(\theta) \\
  & = \sum_{i=1}^n \log \sum_{d_i=0,1} \p(O_i, D_i=d_i|\theta, Y_i)
      \frac{
      \p(D_i=d_i | O_i, \tilde\theta, Y_i)
      }{
      \p(D_i=d_i | O_i, \tilde\theta, Y_i)}
      + \log \p(\theta) \\
  & \geq \sum_{i=1}^n 
            \sum_{d_i=0, 1} 
            \p(D_i=d_i | O_i, \tilde\theta, Y_i) 
            \log \frac{
               \p(O_i, D_i=d_i|\theta, Y_i)
                }{
                \p(D_i=d_i | O_i, \tilde\theta, Y_i)}
        + \log \p(\theta) \\
  & = \sum_{i=1}^n 
            \mathbb{E}_{D_i|O_i, \tilde\theta, Y_i}\Big[
            \log \frac{
               \p(O_i, D_i=d_i|\theta, Y_i)
                }{
                \p(D_i=d_i | O_i, \tilde\theta, Y_i)}\Big]
        + \log \p(\theta) 
    \end{split}
\end{equation}

In the above computations, the lower bound is obtained using Jensen inequality. It represents a tangent lower bound of the posterior distribution: $\theta \rightarrow \ell(\theta)$ at the given set of parameters $\tilde \theta$. On the last line, the expectation is taken for $D_i$ distributed according to $\p(D_i|O_i, \tilde \theta, Y_i)$. Following the literature on the stochastic EM algorithm, we compute this expectation by replacing it with its Monte-Carlo estimate, sampling one $\widehat{D_i}$ according to $\p(D_i|O_i, \tilde  \theta, Y_i)$. For simplicity of notation, we denote:
\begin{equation}
    w_i = \p(D_i=d_i | O_i, \tilde \theta, Y_i).
\end{equation}

The approximate lower bound of the posterior of the parameters writes, after sampling:

\begin{equation}
    \begin{split}
  \ell 
%   \geq \sum_{i=1}^n 
%              \log \frac{
%              \p(T_i, X_i, x, y, s_0, s_1, \beta, D_i  = \widehat{D_i}|Y_i)}{
%              \p(D_i  = \widehat{D_i}| T_i, X_i, Y_i, x, y, s_0, s_1, \beta)}
  \geq \sum_{i=1}^n 
             \log \frac{
             \p(O_i, D_i  = \widehat{D_i}|\theta, Y_i)
             }{
             w_i}
        + \log \p(\theta),
    \end{split}
\end{equation}
and we only need to maximize the following function $\mathcal{M}$ in its parameters $\theta$:
\begin{equation}
    \theta \rightarrow \mathcal{M}(\theta) = \sum_{i=1}^n 
             \log 
             \p(O_i, D_i  = \widehat{D_i}|\theta, Y_i)
        + \log \p(\theta).
\end{equation}

Since $D_i = \widehat{D_i}$ is now fixed, we further decompose the right-hand side of the above inequality.
\begin{equation}
    \begin{split}
  \mathcal{M}(\theta)  
  & = \sum_{i=1}^n 
            \log \p(T_i, S_i, X_i, D_i  = \widehat{D_i}|\theta, Y_i) 
      + \log \p(\theta)\\
  & = \sum_{i=1}^n 
            \log \p(T_i, S_i, X_i | D_i  = \widehat{D_i}, \theta, Y_i) \\
  & \quad + \sum_{i=1}^n \log \p(D_i  = \widehat{D_i}|\theta, Y_i)
          + \log \p(\theta)\\
  & = \sum_{i=1}^n 
            \log \p(T_i| D_i  = \widehat{D_i}, \theta, Y_i)
            + \log \p(S_i, X_i| D_i  = \widehat{D_i}, \theta, Y_i) \\
  & \quad + \sum_{i=1}^n \log \p(D_i  = \widehat{D_i}|\theta, Y_i)
          + \log \p(\theta)\\
  & = \sum_{i=1}^n 
            \log \p(T_i| D_i  = \widehat{D_i}, \theta, Y_i)
            + \log \p(S_i, X_i| D_i  = \widehat{D_i}, \theta, Y_i) \\
  & \quad + \sum_{i=1}^n \log \p(D_i  = \widehat{D_i}|\theta, Y_i)
          + \log \p(\theta)\\
  & = \sum_{i=1}^n 
            \log \p(T_i| D_i  = \widehat{D_i}, \theta, Y_i)
            + \log \p(X_i| S_i, D_i  = \widehat{D_i}, \theta, Y_i)
            + \log \p(S_i | D_i  = \widehat{D_i}, \theta, Y_i)\\
  & \quad + \sum_{i=1}^n \log \p(D_i  = \widehat{D_i}|\theta, Y_i)
          + \log \p(\theta)\\
    \end{split}
\end{equation}
using the conditional independence of $(S_i, X_i)$ and $T_i$ given $D_i$. Using the dependency structure of the variables relying on the Bayesian network from Figure~\ref{fig:model}, we get:
\begin{equation}
    \begin{split}
  \mathcal{M}(\theta)  
  & = \sum_{i=1}^n 
            \log \p(T_i| D_i  = \widehat{D_i}, x, y)\\
  & \quad + \sum_{i=1}^n \log \p(X_i| S_i, D_i  = \widehat{D_i}, s_0, s_1)
          + \log \p(S_i | D_i  = \widehat{D_i}, p_0, p_1)\\
  & \quad + \sum_{i=1}^n \log \p(D_i  = \widehat{D_i}|\beta, Y_i) \\
  & \quad + \log \p(x, y) + \log \p(p_0, p_1) + \log \p(s_0, s_1) + \log \p(\beta)\\
    \end{split}
\end{equation}

As a result, we find the maximum a posteriori of $x,y$, $s_0, s_1$ and $\beta$ separately, by maximizing respectively:
\begin{equation}\label{eq:m}
    \begin{split}
        M(x, y) 
        &= \sum_{i=1}^n 
            \log \p(T_i| D_i  = \widehat{D_i}, x, y) 
            + \log \p(x, y), \\
        M(p_0, p_1) 
        &= \sum_{i=1}^n 
            \log \p(S_i | D_i  = \widehat{D_i}, p_0, p_1) 
            + \log \p(p_0, p_1) \\
        M(s_0, s_1) 
        &= \sum_{i=1}^n 
            \log \p(X_i| S_i, D_i  = \widehat{D_i}, s_0, s_1) 
            + \log \p(s_0, s_1), \\
        M(\beta) & = \sum_{i=1}^n  
            \log \p(D_i=\widehat{D_i}|\beta, Y_i )
            + \log \p(\beta).
    \end{split}
\end{equation}

To summarize, at iteration $(j+1)$ of the stochastic EM algorithm:
\begin{itemize}
    \item Stochastic E-step: For each patient $i$:
    \begin{itemize}
        \item Compute the posterior distribution of their diagnosis $\p(D_i|O_i, \tilde \theta, Y_i)$,
        \item Sample $\widehat{D_i}$ from the posterior.
    \end{itemize}
    \item M-step: Maximize the approximate lower-bound of the parameters' posteriors in $\theta=(x, y, p_0, p_1, s_0, s_1, \beta)$ by maximizing $M(x, y), M(p_0, p_1), M(s_0, s_1)$ and $M(\beta)$. This updates the parameters to:
    \begin{equation}
        \theta^{(j+1)} =( x^{(j+1)}, y^{(j+1)}, p_0^{(j+1)}, p_1^{(j+1)}, s_0^{(j+1)}, s_1^{(j+1)}, \beta^{(j+1)}).
    \end{equation}
\end{itemize}
until convergence in $\theta = (x, y, p_0, p_1, s_0, s_1, \beta)$. At each iteration, this algorithm maximizes a (stochastic approximation) of a tangent lower-bound of the posterior distribution of the parameters. Therefore, each iteration increases the posterior distribution of the parameters.

\subsection{Auxiliary computation: joint log-likelihood}

As an auxiliary computation, we provide the formula for the joint log-likelihood under our model, which we will use in the E- and M-steps.

The joint log-likelihood writes:
\begin{equation}
    \begin{split}
        \p(O, D, \theta|Y) 
        & = \p(T, S, X, D, x, y, p_0, p_1, s_0, s_1, \beta|Y) \\
        & = \p(T, S, X, x, y, p_0, p_1, s_0, s_1| D, \beta, Y) \times \p(D, \beta|Y) \\
        & = \p(T, S, X, x, y, p_0, p_1, s_0, s_1| D) \times \p(D|Y, \beta) \times \p(\beta|Y) \\
      \end{split}
\end{equation}           
Using the conditional independence of $(S, X)$ and $T$ given $D$, we write:
\begin{equation}
    \begin{split}
        & \p(O, D, \theta|Y) \\
        & \quad = \p(T, x, y| D) \times \p(S, X, p_0, p_1, s_0, s_1| D) \times \p(D|Y, \beta) \times \p(\beta|Y) \\
        & \quad = \p(T, x, y| D) \times \p(X, s_0, s_1| S, p_0, p_1, D) \times \p(S, p_0, p_1| D) \times \p(D|Y, \beta) \times \p(\beta|Y) \\
        & \quad = \p(T, x, y| D) \times \p(X, s_0, s_1| S, D) \times \p(S, p_0, p_1| D) \times \p(D|Y, \beta) \times \p(\beta|Y)
    \end{split}
\end{equation}  

\subsubsection{Auxiliary computations using the generative model}\label{sub:aux}

We separately compute the probabilities in the above formula, using the probability distributions provided by the generative model. We get, for the term involving the immunoassay test $T$:
\begin{equation}
    \begin{split}
        \p(T, x, y| D)
        & = \p(T| D, x, y) \times \p(x, y | D) \\
        & = \p(T| D, x, y) \times \p(x) \times \p(y) \\
        & = x^{TD}(1-x)^{(1-T)D}y^{T(1-D)}(1-y)^{(1-T)(1-D)} \\
        & \quad \times B(\alpha_x, \beta_x)^{-1}   
                           x^{\alpha_x - 1} (1-x)^{\beta_x -1}
                \times B(\alpha_y, \beta_y)^{-1} 
                           y^{\alpha_y - 1} (1-y)^{\beta_y -1}.
    \end{split}
\end{equation}
For the term involving the symptoms $X$, we get:
\begin{equation}
    \begin{split}
        \p(X, s_0, s_1| S, D)
        & = \p(X| S, D, s_0, s_1) \times \p(s_0, s_1 |D ) \\
        & = \p(X| S, D, s_0, s_1) \times \p(s_0) \times \p(s_1) \\
        & = 0^{(1-S)X} 1^{(1-S)(1-X)} 
            s_0^{S(1-D)X} (1-s_0)^{S(1-D)(1-X)}
            s_1^{SDX} (1-s_1)^{SD(1-X)} \\
        & \quad \times B(\alpha_{s_0}, \beta_{s_0})^{-1}
                s_0^{\alpha_{s_0} - 1} (1-s_0)^{\beta_{s_0} -1}
          \times B(\alpha_{s_1}, \beta_{s_1})^{-1} 
                s_1^{\alpha_{s_1} - 1} (1-s_1)^{\beta_{s_1} -1} \\
        & = \delta_{S=0, X=0} \\
        & \quad + \delta_{S=1} \times s_0^{(1-D)X} (1-s_0)^{(1-D)(1-X)}
            s_1^{DX} (1-s_1)^{D(1-X)} \\
        & \qquad \quad \quad \times B(\alpha_{s_0}, \beta_{s_0})^{-1}
                s_0^{\alpha_{s_0} - 1} (1-s_0)^{\beta_{s_0} -1} \\
        & \qquad \quad \quad \times B(\alpha_{s_1}, \beta_{s_1})^{-1} 
                s_1^{\alpha_{s_1} - 1} (1-s_1)^{\beta_{s_1} -1}
    \end{split}
\end{equation}
For the term involving the symptomatic variable $S$, we get:
\begin{equation}
    \begin{split}
        \p(S, p_0, p_1| D)
        & = \p(S | p_0, p_1, D) \times \p(p_0) \times \p(p_1) \\
        & = p_0^{(1-D)S} (1-p_0)^{(1-D)(1-S)}
        \times p_1^{DS} (1-p_1)^{D(1-S)} \\
        & \quad \times B(\alpha_{p_0}, \beta_{p_0})^{-1}
                p_0^{\alpha_{p_0} - 1} (1-p_0)^{\beta_{p_0} -1} \\
        & \quad \times B(\alpha_{p_1}, \beta_{p_1})^{-1}
                p_1^{\alpha_{p_1} - 1} (1-p_1)^{\beta_{p_1} -1}.
    \end{split}
\end{equation}
For the terms involving the diagnosis $D$ and the parameter $\beta$ of the logistic regression, we get:
\begin{equation}
    \begin{split}
        \p(D|Y, \beta)
        & =  \pi(Y, \beta)^D (1-\pi(Y, \beta))^{1-D} \\
        \p(\beta) 
        & = n(\beta; \sigma_\beta^2),   \\
    \end{split}
\end{equation}
where we use the notations:
\begin{itemize}
    \item $\pi(Y, \beta)=g(Y\beta)$ with $g$ the sigmoid function from the logistic regression,
    \item $n(\beta; \sigma_\beta^2)$ the probability density function of the Gaussian $\mathcal{N}(0, \sigma_\beta^2)$.
\end{itemize}

\subsubsection{Final formula for the joint log-likelihood}

We plug the expressions of the probabilities in the joint log-likelihood, and get:
 \begin{equation}\label{eq:jll}
    \begin{split}
        & \p(O, D, \theta|Y) \\
        & \quad 
        = 
        \p(T, x, y| D) 
        \times \p(X, s_0, s_1| S, D) 
        \times \p(S, p_0, p_1| D) 
        \times \p(D|Y, \beta) 
        \times \p(\beta|Y)  \\
        & \quad 
        \propto 
        x^{TD}(1-x)^{(1-T)D}y^{T(1-D)}(1-y)^{(1-T)(1-D)} 
        x^{\alpha_x - 1} (1-x)^{\beta_x -1}
        y^{\alpha_y - 1} (1-y)^{\beta_y -1} \\
        & \quad \times 
            0^{(1-S)X} 1^{(1-S)(1-X)} s_0^{S(1-D)X} (1-s_0)^{S(1-D)(1-X)} s_1^{SDX} (1-s_1)^{SD(1-X)}\\
        & \qquad \qquad \times s_0^{\alpha_{s_0} - 1} (1-s_0)^{\beta_{s_0} -1} 
                               s_1^{\alpha_{s_1} - 1} (1-s_1)^{\beta_{s_1} -1}\\
        & \quad \times 
             p_0^{(1-D)S} (1-p_0)^{(1-D)(1-S)} \times p_1^{DS} (1-p_1)^{D(1-S)} \\
        & \qquad \qquad \times p_0^{\alpha_{p_0} - 1} (1-p_0)^{\beta_{p_0} -1} 
                               p_1^{\alpha_{p_1} - 1} (1-p_1)^{\beta_{p_1} -1} \\
        & \quad \times \pi(Y, \beta)^D (1-\pi(Y, \beta))^{1-D} \times n(\beta; \sigma_\beta^2),
    \end{split}
\end{equation}
where we omit the normalizing constants from the beta distributions.

\subsection{Stochastic E-step: Compute posterior of the hidden variables $D_i$}

In the stochastic E-step, we aim to sample $\widehat{D_i}$ from the posterior distribution of the hidden variable $D_i$. Given the current estimate $\theta^{(j)}$ of the parameters, we compute the posterior of the diagnosis $D_i$ for each patient $i$:
\begin{equation}
    \p(D_i|O_i, \theta^{(j)}, Y_i) 
    \propto \p(D_i, O_i, \theta^{(j)} | Y_i),
\end{equation}
where we plug the expression of the joint log-likelihood from Equation~\ref{eq:jll}.

Omitting the coefficients that are shared in both formulae for the probability below, we have:
\begin{equation}
    \begin{split}
        &\p(D_i=1|O_i, \theta^{(j)}, Y_i) \\
        &\quad\propto {x^{(j)}}^{T_i}(1-x^{(j)})^{(1-T_i)}\times {s_1^{(j)}}^{S_iX_i}(1-s_1^{(j)})^{S_i(1-X_i)} \times \pi(Y_i, \beta^{(j)})
        \times p_1^{S_i}(1-p_1)^{(1-S_i)}\\
        &\p(D_i=0|O_i, \theta^{(j)}, Y_i) \\
        &\quad\propto {y^{(j)}}^{T_i}(1-y^{(j)})^{(1-T_i)}\times {s_0^{(j)}}^{S_iX_i}(1-s_0^{(j)})^{S_i(1-X_i)} \times (1-\pi(Y_i, \beta^{(j)}))
        \times p_0^{S}(1-p_0)^{(1-S_i)},\\
    \end{split}
\end{equation}
by identification in the expression of the joint log-likehood from Equation~\ref{eq:jll}. 

This shows the proposition:
\begin{proposition}\label{prop:d}
The odds of the posterior of the hidden variable $D_i$ at iteration $(j+1)$ writes:
\begin{equation}
   \begin{split}
    &\frac{
        \p(D_i=1|T_i, S_i, X_i, Y_i, \theta^{(j)})
        }{
        \p(D_i=0|T_i, S_i, X_i, Y_i, \theta^{(j)})}\\
    &= \frac{
    {x^{(j)}}^{T_i} (1-x^{(j)})^{(1-T_i)}
    \times {s_1^{(j)}}^{S_iX_i} (1-s_1^{(j)})^{S_i(1-X_i)} 
    \times \pi(Y_i, \beta^{(j)})
    \times p_1^{S_i}(1-p_1)^{(1-S_i)}
    }{
    {y^{(j)}}^{T_i}(1-y^{(j)})^{(1-T_i)}\times {s_0^{(j)}}^{S_iX_i}(1-s_0^{(j)})^{S_i(1-X_i)} \times (1-\pi(Y_i, \beta^{(j)}))
        \times p_0^{S_i}(1-p_0)^{(1-S_i)}}.
    \end{split}
\end{equation}
\end{proposition}

Following the methodology of the stochast EM algorithm, we sample from this posterior to obtain $\widehat{D_i}$.

% Since we do not observe the $D_i$, we propose an Expectation-Maximization framework,  filling the missing variables by their imputed scores $\hat{D}_i = \mathbb{P}_{\theta^b}(D=1|Y_i, X_i, T_i)$ (as in Eq.\ref{eq:eq}.
% We then update at the next iterations the score (based on gradient descent for the simple logistic regression model):
% $$\frac{\partial l(\theta)}{\partial \beta_j}  = [\hat{D}_i - Y\beta]\beta_j$$

\subsection{M-step: Update model's parameters}

In the M-step, we update the parameters of the generative model by maximizing the lower-bound of their posterior distribution. We update each set of parameters separetely, using the expressions in Equation~\ref{eq:m}.

\subsubsection{Update Sensitivity and Specificity}

Given $\widehat{D_i}$'s, we update $x, y$ by maximizing:
\begin{equation}
    M(x, y) = \sum_{i=1}^n 
              \log \p(T_i| D_i  = \widehat{D_i}, x, y) 
              + \log \p(x, y).
\end{equation}

We recognize in this expression the logarithm of the probability distribution of a product of beta distributions in $x$ and in $y$, omitting the normalization constants:
\begin{equation}
    \begin{split}
       \Pi_{i=1}^n \p(T_i| D_i  = \widehat{D_i}, x, y)
         & = \Pi_{i=1}^n \Big( 
              x^{T_i \widehat{D_i}}(1-x)^{(1-T_i)\widehat{D_i}} \times y^{T_i(1-\widehat{D_i})}(1-y)^{(1-T_i)(1-\widehat{D_i})} \Big)\\
        \p(x, y) 
         & =  x^{\alpha_x - 1} (1-x)^{\beta_x -1}
           \times y^{\alpha_y - 1} (1-y)^{\beta_y -1},
    \end{split}
\end{equation}
such that $M(x, y)$ writes:
\begin{equation}
\begin{split}
    M(x, y)
    & = \log\Big[
        \mathcal{B}_x \Big(
            \sum_{i=1}^nT_i\widehat{D_i} + \alpha_x, 
            \sum_{i=1}^n(1-T_i)\widehat{D_i} + \beta_x\Big) \Big]\\
    & + \log \Big[ 
        \mathcal{B}_y \Big( 
            \sum_{i=1}^n T_i (1 - \widehat{D_i}) + \alpha_y, 
            \sum_{i=1}^n (1-T_i) (1 - \widehat{D_i}) + \beta_y \Big)\Big].
    \end{split}
\end{equation}

The updated sensitivity and specificity, $x^{(j+1)},y^{(j+1)}$, maximize $M(x, y)$. They are the modes of the beta distributions:
\begin{equation}
    \begin{split}
        x^{(j+1)} &= \underset{x}{\text{argmax }} \mathcal{B}_x\Big(
        \sum_{i=1}^n T_i\widehat{D_i} + \alpha_x, 
        \sum_{i=1}^n (1-T_i)\widehat{D_i} + \beta_x\Big) \\
        y^{(j+1)} & = \underset{y}{\text{argmax }} \mathcal{B}_y\Big(
        \sum_{i=1}^n T_i (1-\widehat{D_i}) + \alpha_y, 
        \sum_{i=1}^n (1-T_i)(1-\widehat{D_i}) + \beta_y \Big).
    \end{split}
\end{equation}

If $\alpha_x, \beta_x > 1 $ and $\alpha_y, \beta_y > 1 $, the expression for the modes are:
\begin{equation}
    \begin{split}
       x^{(j+1)}  
       &= \frac{
       \sum_{i=1}^nT_i\widehat{D_i}+\alpha_x - 1}{
       \sum_{i=1}^nT_i\widehat{D_i}+\alpha_x + \sum_{i=1}^n(1-T_i)\widehat{D_i}+\beta_x -2 } \\
       &= \frac{
       \sum_{i=1}^nT_i\widehat{D_i}+\alpha_x - 1}{
       \sum_{i=1}^n\widehat{D_i}+(\alpha_x+\beta_x) - 2 }\\
       y^{(j+1)}
       &= \frac{
       \sum_{i=1}^nT_i(1-\widehat{D_i})+\alpha_y - 1}{
       \sum_{i=1}^nT_i(1-\widehat{D_i})+\alpha_y+ \sum_{i=1}^n(1-T_i)(1-\widehat{D_i})+\beta_y - 2}\\
       &= \frac{
       \sum_{i=1}^nT_i(1-\widehat{D_i})+\alpha_y - 1}{
       \sum_{i=1}^n (1-\widehat{D_i})+(\alpha_y+\beta_y) - 2}.
    \end{split}
\end{equation}

We can have the case $\beta_x, \beta_y < 1$ if the sensitivity or the specificity are very close to 1. In this case, we update the parameters using the expectation of the Beta distribution, instead of the mode.

\subsubsection{Update Probabilities of being symptomatic}

Given the $\widehat{D_i}$'s, we update $p_0, p_1$ by maximizing:
\begin{equation}
    M(p_0, s_1) = \sum_{i=1}^n 
              \log \p(S_i | D_i  = \widehat{D_i}, p_0, p_1) 
              + \log \p(p_0, p_1) .
\end{equation}

We recognize in this expression the logarithm of the probability distribution of a product of beta distributions in $p_0$ and in $p_1$, omitting the normalization constants :
\begin{equation}
    \begin{split}
       \Pi_{i=1}^n \p(S_i| D_i  = \widehat{D_i}, p_0, p_1)
         & = \Pi_{i=1}^n \Big(
             p_0^{(1-\widehat{D_i})S_i} (1-p_0)^{(1-\widehat{D_i})(1-S_i)}
             \times p_1^{\widehat{D_i}S_i} (1-p_1)^{\widehat{D_i}(1-S_i)}
             \Big) \\
        \p(p_0, p_1) 
         & = p_0^{\alpha_{p_0} - 1} (1-p_0)^{\beta_{p_0} -1}
             \times  p_1^{\alpha_{p_1} - 1} (1-p_1)^{\beta_{p_1} -1}
    \end{split}
\end{equation}
such that $M(p_0, p_1)$ writes:
\begin{equation}
\begin{split}
    M(p_0, p_1)
    & = \log\Big[
        \mathcal{B}_{p_0} \Big(
            \sum_{i=1}^n (1-\widehat{D_i}) S_i + \alpha_{p_0}, 
            \sum_{i=1}^n (1-\widehat{D_i})(1-S_i) + \beta_{p_0}\Big) \Big]\\
    & + \log \Big[ 
        \mathcal{B}_{p_1} \Big( 
            \sum_{i=1}^n \widehat{D_i}S_i + \alpha_{p_1}, 
            \sum_{i=1}^n \widehat{D_i} (1 - S_i) + \beta_{p_1} \Big)\Big].
    \end{split}
\end{equation}

The updated probabilities of being symptomatic, in absence or presence of the disease, $p_0^{(j+1)}, p_1^{(j+1)}$, maximize $M(p_0, p_1)$. They are the modes of the beta distributions:
\begin{equation}
    \begin{split}
        p_0^{(j+1)} &= \underset{x}{\text{argmax }} \mathcal{B}_x\Big(
        \sum_{i=1}^n (1-\widehat{D_i}) S_i + \alpha_{p_0}, 
        \sum_{i=1}^n (1-\widehat{D_i})(1-S_i) + \beta_{p_0}\Big) \\
        p_1^{(j+1)} & = \underset{y}{\text{argmax }} \mathcal{B}_y\Big(
        \sum_{i=1}^n \widehat{D_i}S_i + \alpha_{p_1}, 
        \sum_{i=1}^n \widehat{D_i} (1 - S_i) + \beta_{p_1} \Big).
    \end{split}
\end{equation}

If $\alpha_{p_0}, \beta_{p_0} > 1 $ and $\alpha_{p_1}, \beta_{p_1} > 1 $, the expression for the modes are:
\begin{equation}
    \begin{split}
       {p_0}^{(j+1)}  
       &= \frac{
       \sum_{i=1}^n (1-\widehat{D_i}) S_i + \alpha_{p_0} - 1}{
       \sum_{i=1}^n (1-\widehat{D_i}) S_i + \alpha_{p_0}  
       + \sum_{i=1}^n (1-\widehat{D_i})(1-S_i) + \beta_{p_0} - 2 } \\
       &= \frac{
       \sum_{i=1}^n (1-\widehat{D_i}) S_i + \alpha_{p_0} - 1}{
       \sum_{i=1}^n (1 - \widehat{D_i})+(\alpha_{p_0}+\beta_{p_0}) - 2 }\\
       {p_1}^{(j+1)}
       &= \frac{
       \sum_{i=1}^n \widehat{D_i}S_i + \alpha_{p_1} - 1}{
       \sum_{i=1}^n \widehat{D_i}S_i + \alpha_{p_1} 
       + \sum_{i=1}^n \widehat{D_i} (1 - S_i) + \beta_{p_1} - 2}\\
       &= \frac{
       \sum_{i=1}^n \widehat{D_i}S_i + \alpha_{p_1} - 1}{
       \sum_{i=1}^n \widehat{D_i}+(\alpha_{p_1}+\beta_{p_1}) - 2}.
    \end{split}
\end{equation}

We can have the case $\beta_{p_0}, \beta_{p_1} < 1$ if the sensitivity or the specificity are very close to 1. In this case, we update the parameters using the expectation of the Beta distribution, instead of the mode.

\subsubsection{Update Symptoms Probabilities}

Given $\widehat{D_i}$'s, we update $s_0, s_1$ by maximizing:
\begin{equation}
    M(s_0, s_1) = \sum_{i=1}^n 
              \log \p(X_i| S_i, D_i  = \widehat{D_i}, s_0, s_1) 
              + \log \p(s_0, s_1), .
\end{equation}

We recognize in this expression the logarithm of the probability distribution of a product of beta distributions in $s_0$ and in $s_1$, omitting the normalization constants:
\begin{equation}
    \begin{split}
        & \Pi_{i=1}^n \p(X_i| S_i, D_i  = \widehat{D_i}, s_0, s_1) \\
        & = \Pi_{i=1}^n \Big( \delta_{S_i=0, X_i=0} 
           + \delta_{S_i=1} \times
                  s_0^{(1-\widehat{D_i})X_i} (1-s_0)^{(1-\widehat{D_i})(1-X_i)}
                  s_1^{\widehat{D_i}X_i} (1-s_1)^{\widehat{D_i}(1-X_i)}
             \Big) \\
        & = \Pi_{i=1, \text{s.t. } S_i=1}^n \Big(
            s_0^{(1-\widehat{D_i})X_i} (1-s_0)^{(1-\widehat{D_i})(1-X_i)}
            s_1^{\widehat{D_i} X_i} (1-s_1)^{\widehat{D_i}(1-X_i)}
             \Big)
    \end{split}
\end{equation}
and:
\begin{equation}
\begin{split}
        \p(s_0, s_1) 
         & = s_0^{\alpha_{s_0} - 1} (1-s_0)^{\beta_{s_0} -1}
             \times  s_1^{\alpha_{s_1} - 1} (1-s_1)^{\beta_{s_1} -1}
    \end{split}
\end{equation}
such that $M(s_0, s_1)$ writes:
\begin{equation}
\begin{split}
    M(s_0, s_1)
    & = \log\Big[
        \mathcal{B}_{s_0} \Big(
            \sum_{i=1, \text{s.t. } S_i=1}^n (1-\widehat{D_i}) X_i + \alpha_{s_0}, 
            \sum_{i=1, \text{s.t. } S_i=1}^n (1-\widehat{D_i})(1-X_i) + \beta_{s_0}\Big) \Big]\\
    & + \log \Big[ 
        \mathcal{B}_{s_1} \Big( 
            \sum_{i=1, \text{s.t. } S_i=1}^n \widehat{D_i}X_i + \alpha_{s_1}, 
            \sum_{i=1, \text{s.t. } S_i=1}^n \widehat{D_i} (1 - X_i) + \beta_{s_1} \Big)\Big].
    \end{split}
\end{equation}

The updated probabilities of exhibiting symptoms, in absence or presence of the disease, $s_0^{(j+1)}, s_1^{(j+1)}$, maximize $M(s_0, s_1)$. They are the modes of the beta distributions:
\begin{equation}
    \begin{split}
        s_0^{(j+1)} &= \underset{x}{\text{argmax }} \mathcal{B}_x\Big(
        \sum_{i=1, \text{s.t. } S_i=1}^n (1-\widehat{D_i}) X_i + \alpha_{s_0}, 
        \sum_{i=1, \text{s.t. } S_i=1}^n (1-\widehat{D_i})(1-X_i) + \beta_{s_0}\Big) \\
        s_1^{(j+1)} & = \underset{y}{\text{argmax }} \mathcal{B}_y\Big(
        \sum_{i=1, \text{s.t. } S_i=1}^n \widehat{D_i}X_i + \alpha_{s_1}, 
        \sum_{i=1, \text{s.t. } S_i=1}^n \widehat{D_i} (1 - X_i) + \beta_{s_1} \Big).
    \end{split}
\end{equation}

If $\alpha_{s_0}, \beta_{s_0} > 1 $ and $\alpha_{s_1}, \beta_{s_1} > 1 $, the expression for the modes are:
\begin{equation}
    \begin{split}
       {s_0}^{(j+1)}  
       &= \frac{
       \sum_{i=1, \text{s.t. } S_i=1}^n (1-\widehat{D_i}) X_i + \alpha_{s_0} - 1}{
       \sum_{i=1, \text{s.t. } S_i=1}^n (1-\widehat{D_i}) X_i + \alpha_{s_0}  
       + \sum_{i=1, \text{s.t. } S_i=1}^n (1-\widehat{D_i})(1-X_i) + \beta_{s_0} - 2 } \\
       &= \frac{
       \sum_{i=1, \text{s.t. } S_i=1}^n (1-\widehat{D_i}) X_i + \alpha_{s_0} - 1}{
       \sum_{i=1, \text{s.t. } S_i=1}^n (1 - \widehat{D_i})+(\alpha_{s_0}+\beta_{s_0}) - 2 }\\
       {s_1}^{(j+1)}
       &= \frac{
       \sum_{i=1, \text{s.t. } S_i=1}^n \widehat{D_i}X + \alpha_{s_1} - 1}{
       \sum_{i=1, \text{s.t. } S_i=1}^n \widehat{D_i}X_i + \alpha_{s_1} 
       + \sum_{i=1, \text{s.t. } S_i=1}^n \widehat{D_i} (1 - X_i) + \beta_{s_1} - 2}\\
       &= \frac{
       \sum_{i=1, \text{s.t. } S_i=1}^n \widehat{D_i}X_i + \alpha_{s_1} - 1}{
       \sum_{i=1, \text{s.t. } S_i=1}^n \widehat{D_i}+(\alpha_{s_1}+\beta_{s_1}) - 2}.
    \end{split}
\end{equation}

We can have the case $\beta_{s_0}, \beta_{s_1} < 1$ if the probabilities of symptoms are very close to 1. In this case, we update the parameters using the expectation of the Beta distribution, instead of the mode.

\subsubsection{Update coefficients of the risk factors $\beta$}

Given the $\widehat{D_i}$'s, we update $\beta$ by maximizing:
\begin{equation}
    \begin{split}
        M(\beta) 
        & = \sum_{i=1}^n  
             \log \p(D_i=\widehat{D_i}|\beta, Y_i)
             + \log \p(\beta)\\
        &= -  \sum_{i=1}^n \frac{||\widehat{D_i} - g(Y_i\beta)||^2}{2\sigma^2} - \frac{||\beta||^2}{2\sigma_\beta^2}
    \end{split}
\end{equation}
where we omit the normalization constant of the Gaussian distributions. We recognize the loss function of the logistic regression, that we optimize using stochastic gradient descent, with update rule:
\begin{equation}
    \beta 
    \leftarrow \beta - \gamma \Big[ 
        \sum_{i} (g(Y_i\beta) - \widehat{D_i})Y_i + \frac{||\beta||^2}{2\sigma_\beta^2}\Big]
\end{equation}

This gives the update in $\beta$.

\subsubsection{Summary of the updates}

This proposition summarizes the parameters updates.

\begin{proposition}\label{prop:xy}
The parameters updates write:
\begin{equation}
    \begin{gathered}
       x^{(j+1)}  
       = \frac{
       \sum_{i=1}^nT_i\widehat{D_i}+\alpha_x - 1}{
       \sum_{i=1}^n\widehat{D_i}+(\alpha_x+\beta_x) - 2 }, \quad
       y^{(j+1)}
       = \frac{
       \sum_{i=1}^nT_i(1-\widehat{D_i})+\alpha_y - 1}{
       \sum_{i=1}^n (1-\widehat{D_i})+(\alpha_y+\beta_y) - 2}\\
       p_0^{(j+1)} 
       = \frac{
       \sum_{i=1}^n (1-\widehat{D_i}) S_i + \alpha_{p_0} - 1}{
       \sum_{i=1}^n (1 - \widehat{D_i})+(\alpha_{p_0}+\beta_{p_0}) - 2 }, \quad
       p_1^{(j+1)} 
        = \frac{
       \sum_{i=1}^n \widehat{D_i}S_i + \alpha_{p_1} - 1}{
       \sum_{i=1}^n \widehat{D_i}+(\alpha_{p_1}+\beta_{p_1}) - 2}\\
       s_0^{(j+1)} 
       = \frac{
       \sum_{i=1, \text{s.t. } S_i=1}^n (1-\widehat{D_i}) X_i + \alpha_{s_0} - 1}{
       \sum_{i=1, \text{s.t. } S_i=1}^n (1 - \widehat{D_i})+(\alpha_{s_0}+\beta_{s_0}) - 2 }, \quad
       s_1^{(j+1)} 
        = \frac{
       \sum_{i=1, \text{s.t. } S_i=1}^n \widehat{D_i}X_i + \alpha_{s_1} - 1}{
       \sum_{i=1, \text{s.t. } S_i=1}^n \widehat{D_i}+(\alpha_{s_1}+\beta_{s_1}) - 2}\\
       \beta^{(j+1)}
        = \text{argmin}_\beta \sum_{i=1}^n \frac{||\widehat{D_i} - g(Y_i\beta)||^2}{2\sigma^2} + \frac{||\beta||^2}{2\sigma_\beta^2},
    \end{gathered}
\end{equation}
where the minimization on $\beta$ is performed through stochastic gradient descent.
\end{proposition}

\section{Stochastic EM with missing $T$: truncated Bayesian network}

We apply the stochastic EM algorithm in the Bayesian model truncated at $T$. In this model:
\begin{itemize}
    \item $\zeta = \Big(\alpha_{p_0}, \beta_{p_0}, \alpha_{p_1}, \beta_{p_1}, \{\alpha_{s_{0k}}, \beta_{s_{0k}}\}_k, \{ \alpha_{s_{1k}}, \beta_{s_{1k}}\}_k, \{\sigma_{\beta_m}\}_m\Big)$ are hyper-parameters, considered fixed and known,
    \item $\theta = \Big(p_0, p_1, \{s_{0k}, s_{1k}\}_k, \{\beta_m\}_m \Big)$ are the parameters,
    \item $D$ is a hidden random variable,
    \item $Y, S, X$ are the observed variables.
\end{itemize}
We now writ: $O = (S, X)$.

We still wish to maximize the expression:
\begin{equation}
    \begin{split}
  \ell 
  & = \sum_{i=1}^n 
            \mathbb{E}_{D_i|O_i, \tilde\theta, Y_i}\Big[
            \log \frac{
               \p(O_i, D_i=d_i|\theta, Y_i)
                }{
                \p(D_i=d_i | O_i, \tilde\theta, Y_i)}\Big]
        + \log \p(\theta),
    \end{split}
\end{equation}
under our new notations stated above. The approximate lower bound of the posterior of the parameters still writes, after sampling one $\widehat{D_i}$ according to $\p(D_i|O_i, \tilde  \theta, Y_i)$:

\begin{equation}
    \begin{split}
  \ell 
%   \geq \sum_{i=1}^n 
%              \log \frac{
%              \p(T_i, X_i, x, y, s_0, s_1, \beta, D_i  = \widehat{D_i}|Y_i)}{
%              \p(D_i  = \widehat{D_i}| T_i, X_i, Y_i, x, y, s_0, s_1, \beta)}
  \geq \sum_{i=1}^n 
             \log \frac{
             \p(O_i, D_i  = \widehat{D_i}|\theta, Y_i)
             }{
             w_i}
        + \log \p(\theta),
    \end{split}
\end{equation}
, using the notation: $w_i = \p(D_i|O_i, \tilde{\theta}, Y_i)$. Again, we only need to maximize the following function $\mathcal{M}$ in its parameters $\theta$:
\begin{equation}
    \theta \rightarrow \mathcal{M}(\theta) = \sum_{i=1}^n 
             \log 
             \p(O_i, D_i  = \widehat{D_i}|\theta, Y_i)
        + \log \p(\theta).
\end{equation}

Since $D_i = \widehat{D_i}$ is now fixed, we further decompose the right-hand side of the above inequality. 
\begin{equation}
    \begin{split}
  \mathcal{M}(\theta)  
  & = \sum_{i=1}^n 
            \log \p(S_i, X_i, D_i  = \widehat{D_i}|\theta, Y_i) 
      + \log \p(\theta)\\
  & = M(p_0, p_1) + M(s_0, s_1) + M(\beta)\\
    \end{split}
\end{equation}
where:
\begin{equation}\label{eq:m}
    \begin{split}
        M(p_0, p_1) 
        &= \sum_{i=1}^n 
            \log \p(S_i | D_i  = \widehat{D_i}, p_0, p_1) 
            + \log \p(p_0, p_1) \\
        M(s_0, s_1) 
        &= \sum_{i=1}^n 
            \log \p(X_i| S_i, D_i  = \widehat{D_i}, s_0, s_1) 
            + \log \p(s_0, s_1), \\
        M(\beta) & = \sum_{i=1}^n  
            \log \p(D_i=\widehat{D_i}|\beta, Y_i )
            + \log \p(\beta).
    \end{split}
\end{equation}
are the same functions as in the case with observed $T_i$. The only difference is that $\hat{D_i}$ is sampled from $P(D_i|O_i, \tilde{\theta}, Y_i)$ where $O_i = (S_i, X_i)$ does not contain $T_i$.

\subsection{Auxiliary computation: log-likelihood in truncated Bayesian model}

The joint log-likelihood in the truncated Bayesian model writes:
\begin{equation}
    \begin{split}
        \p(O, D, \theta|Y) 
        & = \p(S, X, D, x, y, p_0, p_1, s_0, s_1, \beta|Y) \\
        & = \p(S, X, x, y, p_0, p_1, s_0, s_1| D) \times \p(D|Y, \beta) \times \p(\beta|Y) \\
        & = \p(X, s_0, s_1| S, D) \times \p(S, p_0, p_1| D) \times \p(D|Y, \beta) \times \p(\beta|Y),
      \end{split}
\end{equation}           
which gives the same result as in the non-truncated case, but without the term $\p(T, x, y|D)$. We use the computations from subsection~\ref{sub:aux} to get the final expression of the loglikelihood:
We plug the expressions of the probabilities in the joint log-likelihood, and get:
 \begin{equation}\label{eq:jll}
    \begin{split}
        & \p(O, D, \theta|Y) \\
        & \quad 
        \propto 
            0^{(1-S)X} 1^{(1-S)(1-X)} s_0^{S(1-D)X} (1-s_0)^{S(1-D)(1-X)} s_1^{SDX} (1-s_1)^{SD(1-X)}\\
        & \qquad \qquad \times s_0^{\alpha_{s_0} - 1} (1-s_0)^{\beta_{s_0} -1} 
                               s_1^{\alpha_{s_1} - 1} (1-s_1)^{\beta_{s_1} -1}\\
        & \quad \times 
             p_0^{(1-D)S} (1-p_0)^{(1-D)(1-S)} \times p_1^{DS} (1-p_1)^{D(1-S)} \\
        & \qquad \qquad \times p_0^{\alpha_{p_0} - 1} (1-p_0)^{\beta_{p_0} -1} 
                               p_1^{\alpha_{p_1} - 1} (1-p_1)^{\beta_{p_1} -1} \\
        & \quad \times \pi(Y, \beta)^D (1-\pi(Y, \beta))^{1-D} \times n(\beta; \sigma_\beta^2),
    \end{split}
\end{equation}
where we omit the normalizing constants from the beta distributions.

\subsection{Stochastic E-step in the truncated model: Compute posterior of the hidden variables $D_i$}

In the stochastic E-step, we aim to sample $\widehat{D_i}$ from the posterior distribution of the hidden variable $D_i$. Given the current estimate $\theta^{(j)}$ of the parameters, we compute the posterior of the diagnosis $D_i$ for each patient $i$:
\begin{equation}
    \p(D_i|O_i, \theta^{(j)}, Y_i) = \p(D_i|S_i, X_i, \theta^{(j)}, Y_i) 
    \propto \p(D_i, S_i, X_i, \theta^{(j)} | Y_i),
\end{equation}
where we plug the expression of the joint log-likelihood of the truncated model.

Omitting the coefficients that are shared in both formulae for the probability below, we have:
\begin{equation}
    \begin{split}
        &\p(D_i=1|O_i, \theta^{(j)}, Y_i) \\
        &\quad\propto {s_1^{(j)}}^{S_iX_i}(1-s_1^{(j)})^{S_i(1-X_i)} \times \pi(Y_i, \beta^{(j)})
        \times p_1^{S_i}(1-p_1)^{(1-S_i)}\\
        &\p(D_i=0|O_i, \theta^{(j)}, Y_i) \\
        &\quad\propto {s_0^{(j)}}^{S_iX_i}(1-s_0^{(j)})^{S_i(1-X_i)} \times (1-\pi(Y_i, \beta^{(j)}))
        \times p_0^{S}(1-p_0)^{(1-S_i)},\\
    \end{split}
\end{equation}
by identification. 

\section{Stochastic EM with missing $T$: Missing and hidden variables}

We derive the Stochastic EM algorithm in the full Bayesian model, taking into account the missing variables $T_i$'s (missing for $i=r+1...n$) and hidden variables $D_i$'s. We want to maximize the posterior distribution of the parameters $\theta$:
\begin{equation}
\begin{split}
        \p(\theta |O_1, ... O_r, O_{r+1}^M, ..., O_n^M, Y_1, ..., Y_n) 
             & = \frac{\p(O_1, ...,  O_r, O_{r+1}^M, ..., O_n^M|\theta, Y_1, ..., Y_n) \times \p(\theta)
                }{
                \p(O_1, ..., O_r, O_{r+1}^M, ..., O_n^M|Y_1, ..., Y_n)} \\
             & \propto \Pi_{i=1}^r \p(O_i|\theta, Y_i) \times \Pi_{i=r+1}^n \p(O_i^M|\theta, Y_i) \times \p(\theta)
\end{split}
\end{equation}
where we write $O_i = (S_i, X_i, T_i)$ when $T_i$ is available and $O_i^M = (S_i, X_i)$ when $T_i$ is missing.

This translates into maximizing the expression:
\begin{equation}
    \begin{split}
  \ell 
  & = \sum_{i=1}^r \log \p(O_i|\theta, Y_i) + \sum_{i=r+1}^n \log \p(O_i^M|\theta, Y_i) 
      + \log \p(\theta) \\
  & \geq \sum_{i=1}^r 
            \mathbb{E}_{D_i|O_i, \tilde\theta, Y_i}\Big[
            \log \frac{
               \p(O_i, D_i=d_i|\theta, Y_i)
                }{
                \p(D_i=d_i | O_i, \tilde\theta, Y_i)}\Big]\\
  & \quad\quad + \sum_{i=r+1}^n 
            \mathbb{E}_{D_i, T_i|O_i^M, \tilde\theta, Y_i}\Big[
            \log \frac{
               \p(O_i^M, D_i=d_i, T_i=t_i|\theta, Y_i)
                }{
                \p(D_i=d_i, T_i=t_i | O_i^M, \tilde\theta, Y_i)}\Big]
        + \log \p(\theta),
    \end{split}
\end{equation}
where we use Jensen inequality to compute the tangent lower-bounds of $\ell_D = \sum_{i=1}^r \log \p(O_i|\theta, Y_i)  $ and $\ell_{T, D}=\sum_{i=r+1}^n \log \p(O_i^M|\theta, Y_i) $ independently. This is still a valid lower-bound of $\ell$ at $\tilde{\theta}$ as the sum of the tangent-lower bounds is a tangent-lower bound of the sum. 

After sampling, we need to maximize the following function of $\theta$:
\begin{equation}
\begin{split}
    \theta \rightarrow \mathcal{M}^{MIS}(\theta) 
        & = \sum_{i=1}^r 
               \log 
                   \p(O_i, \widehat{D_i} =d_i|\theta, Y_i)
        + \sum_{i=r+1}^n 
            \log 
               \p(O_i^M, \widehat{D_i}=d_i, \widehat{T_i}=t_i|\theta, Y_i)
        + \log \p(\theta) \\
        & = \mathcal{M}(\theta, n=r) + \sum_{i=r+1}^n 
            \log 
               \p(O_i^M, \widehat{D_i}=d_i, \widehat{T_i}=t_i|\theta, Y_i)
\end{split}
\end{equation}
where $\mathcal{M}(\theta, n=r)$ is the cost function in the case without missing $T$.

We compute the second term of the lower-bound, which we denote $\mathcal{M}^{NEW}$:
\begin{equation}
    \begin{split}
        \mathcal{M}^{NEW}(\theta) 
        & = \sum_{i=r+1}^n 
            \log 
               \p(O_i^M, \widehat{D_i}=d_i, \widehat{T_i}=t_i|\theta, Y_i)\\
        & = \sum_{i=r+1}^n 
            \log 
               \p(S_i, X_i, \widehat{D_i}=d_i, \widehat{T_i}=t_i|\theta, Y_i)\\
        & = \sum_{i=r+1}^n 
            \log \p(T_i=\widehat{T_i}| D_i  = \widehat{D_i}, \theta, Y_i)
            + \log \p(X_i| S_i, D_i  = \widehat{D_i}, \theta, Y_i)
            + \log \p(S_i | D_i  = \widehat{D_i}, \theta, Y_i)\\
  & \quad + \sum_{i=1}^n \log \p(D_i  = \widehat{D_i}|\theta, Y_i)
    \end{split}
\end{equation}
which is $\mathcal{M}(\theta, n=(n-r+1), T_i \rightarrow \widehat{T_i})$. Therefore: $mathcal{M}^{MIS}(\theta)$ is $\mathcal{M(\theta)}$ where the missing $T_i$'s have been replaced by theirs imputed value.

We proceed with the Stochastic EM algorithm as follows:
\begin{itemize}
    \item M-step: Compute the tangent-lower bound, which amounts to:\begin{itemize}
        \item for $i=1, ..., r$: sample $\widehat{D_i}$ from $\p(D_i|O_i, \tilde\theta, Y_i)$,
        \item for $i=r+1, ..., n$: sample $\widehat{D_i}, \widehat{T_i}$ from $\p(D_i, T_i|O_i^M, \tilde\theta, Y_i)$,
    \end{itemize}
    \item E-step: Maximize the lower-bound in $\theta$.
\end{itemize}

\subsection{M-step}

For $i=1, ..., r$, sampling $\widehat{D_i}$ is performed as usual. We focus on sampling $\widehat{D_i}, \widehat{T_i}$ from $\p(D_i, T_i|O_i^M, \tilde\theta, Y_i)$.

We compute the posterior of interest:
\begin{equation}
    \p(D_i, T_i|O_i^M, \tilde\theta, Y_i) \propto \p(D_i, T_i, O_i^M, |\tilde\theta, Y_i) 
\end{equation}

We get:
\begin{equation}
\begin{split}
    & \p(D_i=0, T_i=0|O_i^M, \tilde\theta, Y_i) \\
    & \quad\quad\quad \propto 
    (1-y^{(j)})\times {s_0^{(j)}}^{S_iX_i}(1-s_0^{(j)})^{S_i(1-X_i)} \times (1-\pi(Y_i, \beta^{(j)}))
        \times p_0^{S}(1-p_0)^{(1-S_i)}\\
    & \p(D_i=0, T_i=1|O_i^M, \tilde\theta, Y_i) \\
    & \quad\quad\quad  \propto {y^{(j)}}\times {s_0^{(j)}}^{S_iX_i}(1-s_0^{(j)})^{S_i(1-X_i)} \times (1-\pi(Y_i, \beta^{(j)}))
        \times p_0^{S}(1-p_0)^{(1-S_i)} \\
    & \p(D_i=1, T_i=0|O_i^M, \tilde\theta, Y_i) \\
    & \quad\quad\quad  \propto (1-x^{(j)})\times {s_1^{(j)}}^{S_iX_i}(1-s_1^{(j)})^{S_i(1-X_i)} \times \pi(Y_i, \beta^{(j)})
        \times p_1^{S_i}(1-p_1)^{(1-S_i)}\\
    & \p(D_i=1, T_i=1|O_i^M, \tilde\theta, Y_i) \\
    & \quad\quad\quad  \propto {x^{(j)}}\times {s_1^{(j)}}^{S_iX_i}(1-s_1^{(j)})^{S_i(1-X_i)} \times \pi(Y_i, \beta^{(j)})
        \times p_1^{S_i}(1-p_1)^{(1-S_i)}.
\end{split}
\end{equation}

These allow to sample $(\widehat{D_i}, \widehat{T_i})$ according to the posterior, for $i=r+1, .., n$.

\subsection{E-step}

The update are the same, except that the missing $T_i$s are replaced by their imputed values.

\subsection{Enhancement}

In addition, we can consider that we have a different prior on the imputed $T_i$'s. Therefore, we create a new class of sensivitivy/specificity pair, designed for the imputed $T_i$. The priors are initialized as non-information Beta distribution with parameters $(2, 2)$, and updated at training time.

\section{Additional Plots for Validation on Synthetic Data}

This section provides additional details on the validation of the StEM algorithm on synthetic data.

\xhdr{Improvement upon $T$} Table~\ref{tab:multicol} shows the improvement in the diagnosis' accuracy provided by our method against the sole test $T$, and highlights the potential strength of harvesting multiple noisy sources of information. The values that we have chosen here for the sensitivity are reflective of the ones that have been reported for the LFA test in the COV-CLEAR study.

\begin{table}[ht]
\caption{Gain in accuracy (mean and standard deviation) when using StEM over the sole test}
\begin{center}
\begin{tabular}{|c|c|cccc|}
    \hline
    \multicolumn{2}{|c|}{Sensitivity}& \multicolumn{4}{|c|}{Specificity}\\\hline \hline
   Real-life & Value&  70 & 80 & 93 & 99\\ \hline 
    Asymptomatic & 70 & 16.2 $\pm 3.9$ & 12.8 $\pm 4.0$ & 8.42 $\pm 3.2$&  5.64 $\pm 3.2$\\ \hline
    2-10 days & 80 & 12.3 $\pm 3.2$ & 9.27 $\pm 3.1$& 4.82$\pm 2.9$&97.0 $\pm 2.3$\\ \hline
    11-20 days & 93 & 8.5 $\pm 2.6$ & 6.4 $\pm 2.5$& 2.3 $\pm 2.1$&0.03 $\pm 1.4$\\ \hline
    21+ days & 99.0  & 8.7 $\pm 2.7$ & 5.7 $\pm 2.0$& 2.0 $\pm 1.4$& 0.01 $\pm 0.0$\\ \hline
\end{tabular}
\end{center}
\label{tab:multicol}
\end{table}

\xhdr{Benchmarks} We complete here our discussion of the improvement that our method brings upon Computer-Aided Diagnosis (CAD) standard methods. Fig.~\ref{fig:heatmap_acc} compares --- for various pairs of sensitivity and specificity. --- the diagnosis accuracies of the Stochastic EM algorithm (StEM) and two variants of the EM algorithm: one variant where the parameters' priors are agnostic, or uninformative; and another variant where these priors are computed from the available data, as described in Section 5. This figure demonstrates that learning the parameters' posterior distributions while estimating the diagnosis yields higher prediction accuracy.

\begin{figure}
\centering
    \includegraphics[width=\textwidth]{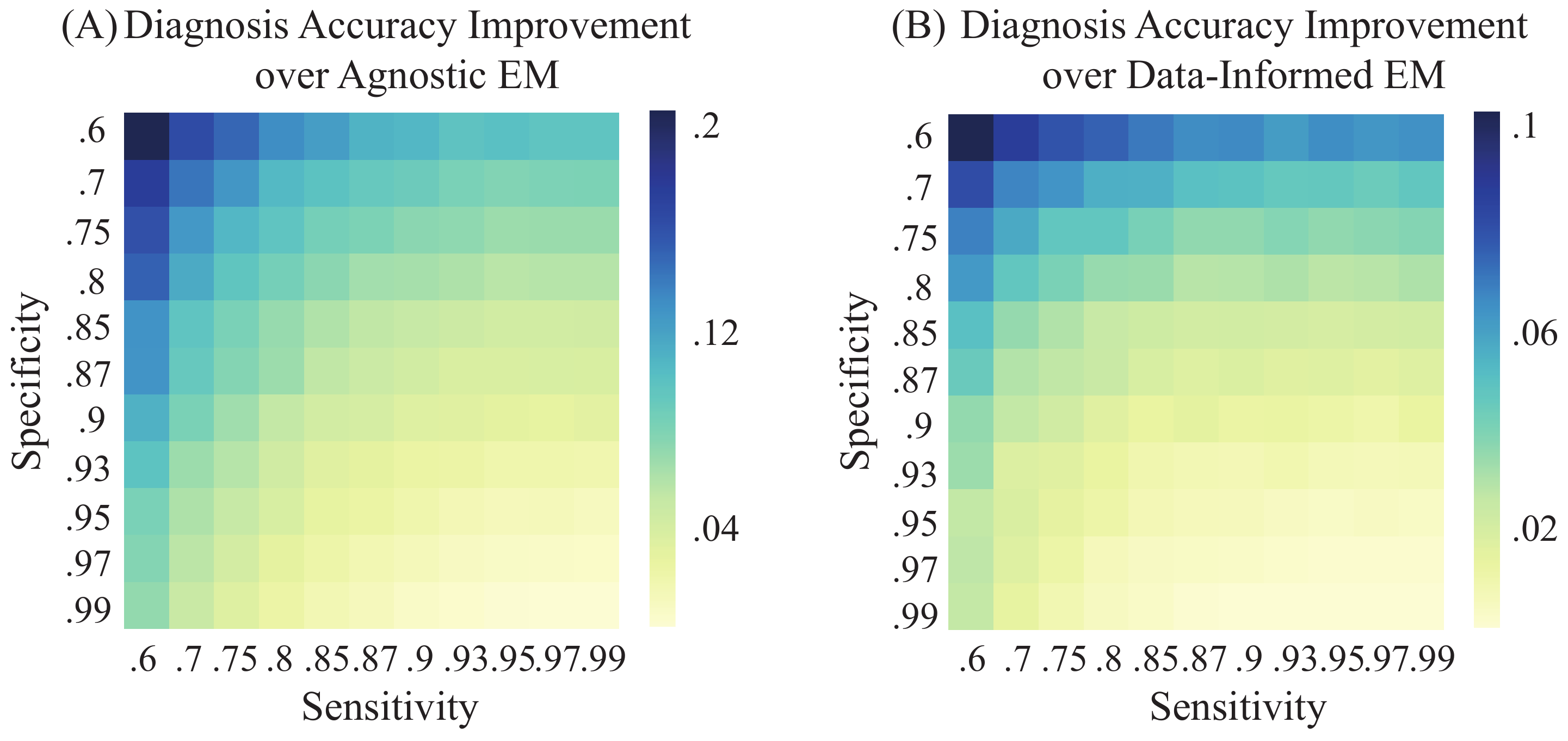}
    \caption{Performance of the StEM algorithm compared to benchmark versions of the EM for n=300 samples, $\sigma=0.5$, and varying levels of specificity and sensitivity. \textbf{(A)} Gain in accuracy with respect to the Data-Agnostic EM described in Section 5. \textbf{(B)} Gain in accuracy with respect to the Data-Informed EM described in Section 5.}
    \label{fig:heatmap_acc}
\end{figure}

\begin{figure}
    \centering
    \includegraphics[width=\textwidth]{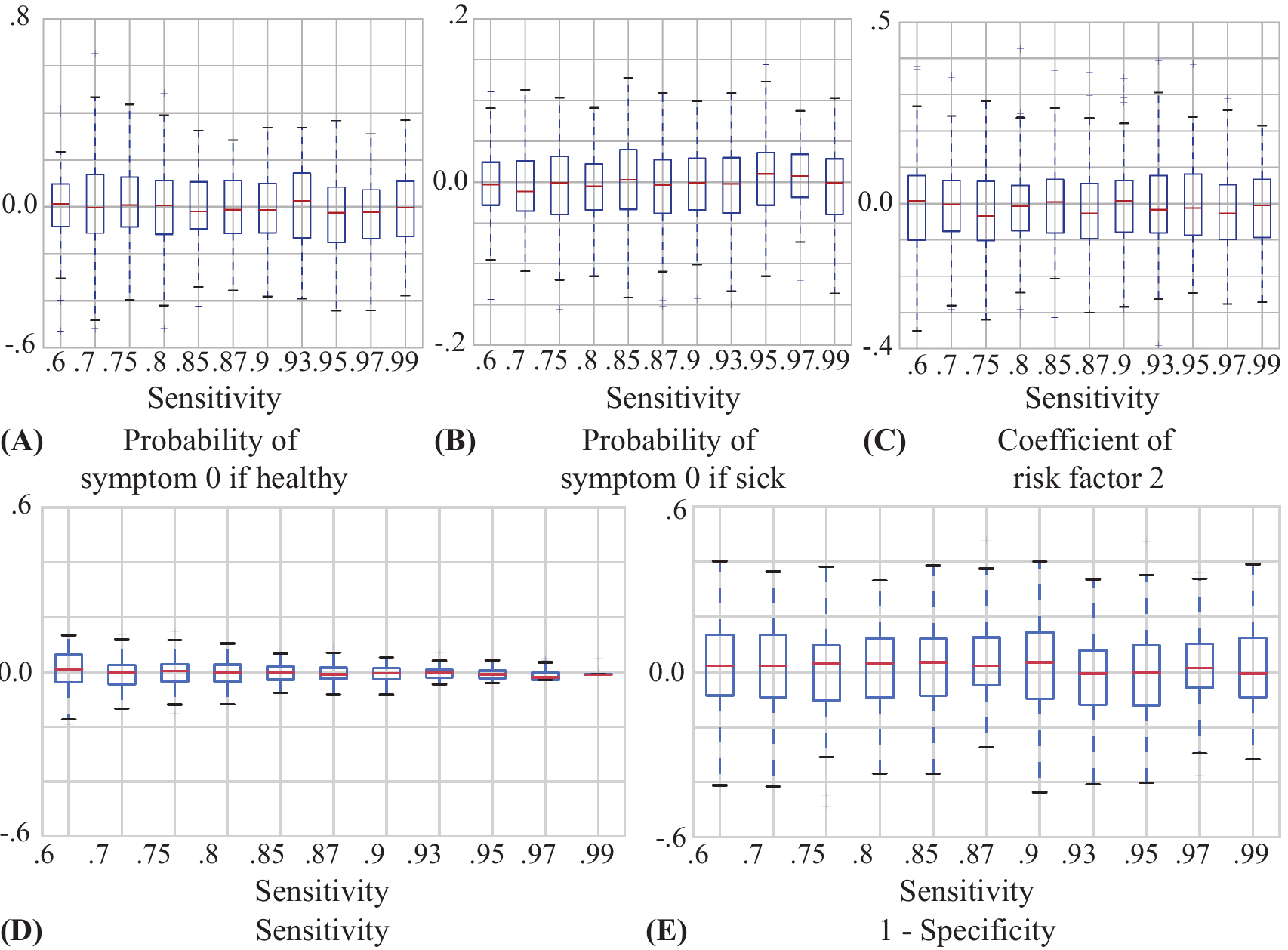}
    \caption{Relative difference between estimated parameters and their ground truth, for $n=300$, $\sigma = 0.5$ and different values of sensitivity (simulated data).}
    \label{fig:results_diff}
\end{figure}

\xhdr{Convergence} Fig.~\ref{fig:results_diff} shows the distribution of the relative difference between recovered coefficients and ground truth (as a percentage of the ground truth value), showing deviations that are within a few percentages of the true value of the coefficients --- thus highlighting the ability of the model to converge to the ground truth parameters.

\begin{figure}
    \centering
    \includegraphics[width=\textwidth]{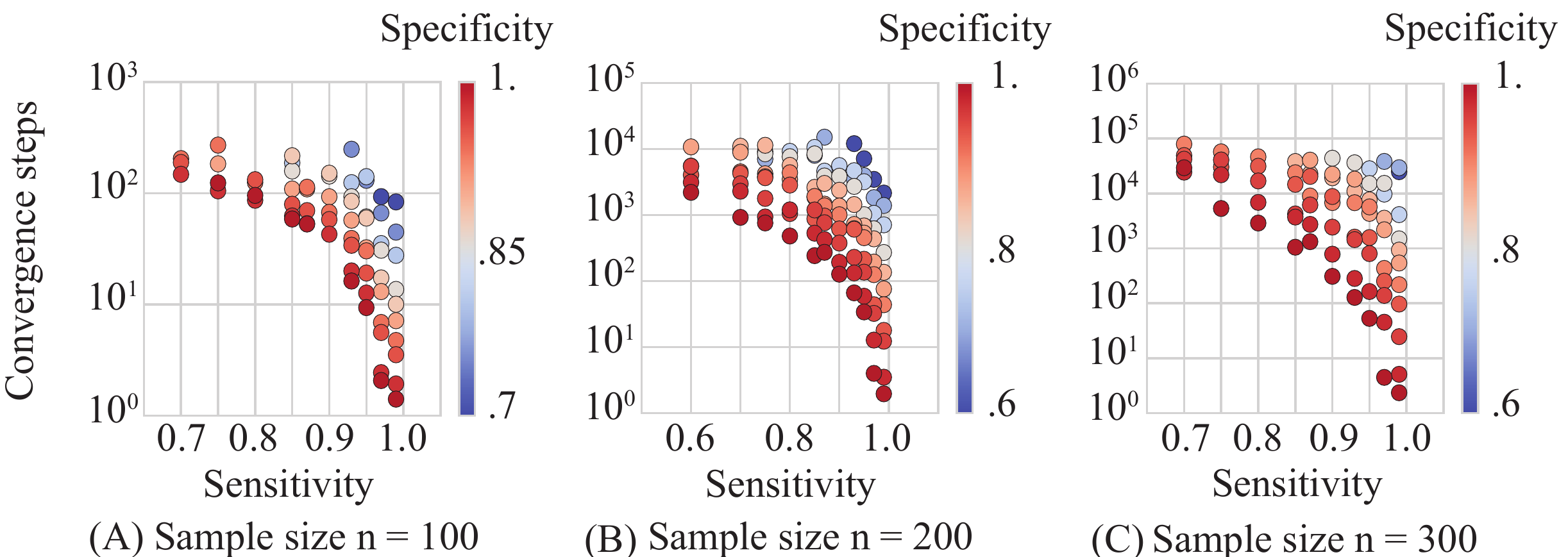}
    \caption{Number of steps until convergence for $\sigma = 0.5$ and different values of sensitivity, specificity and sample sizes (simulated data).}
    \label{fig:results_conv}
\end{figure}

\begin{figure}
    \centering
    \includegraphics[width=0.5\textwidth]{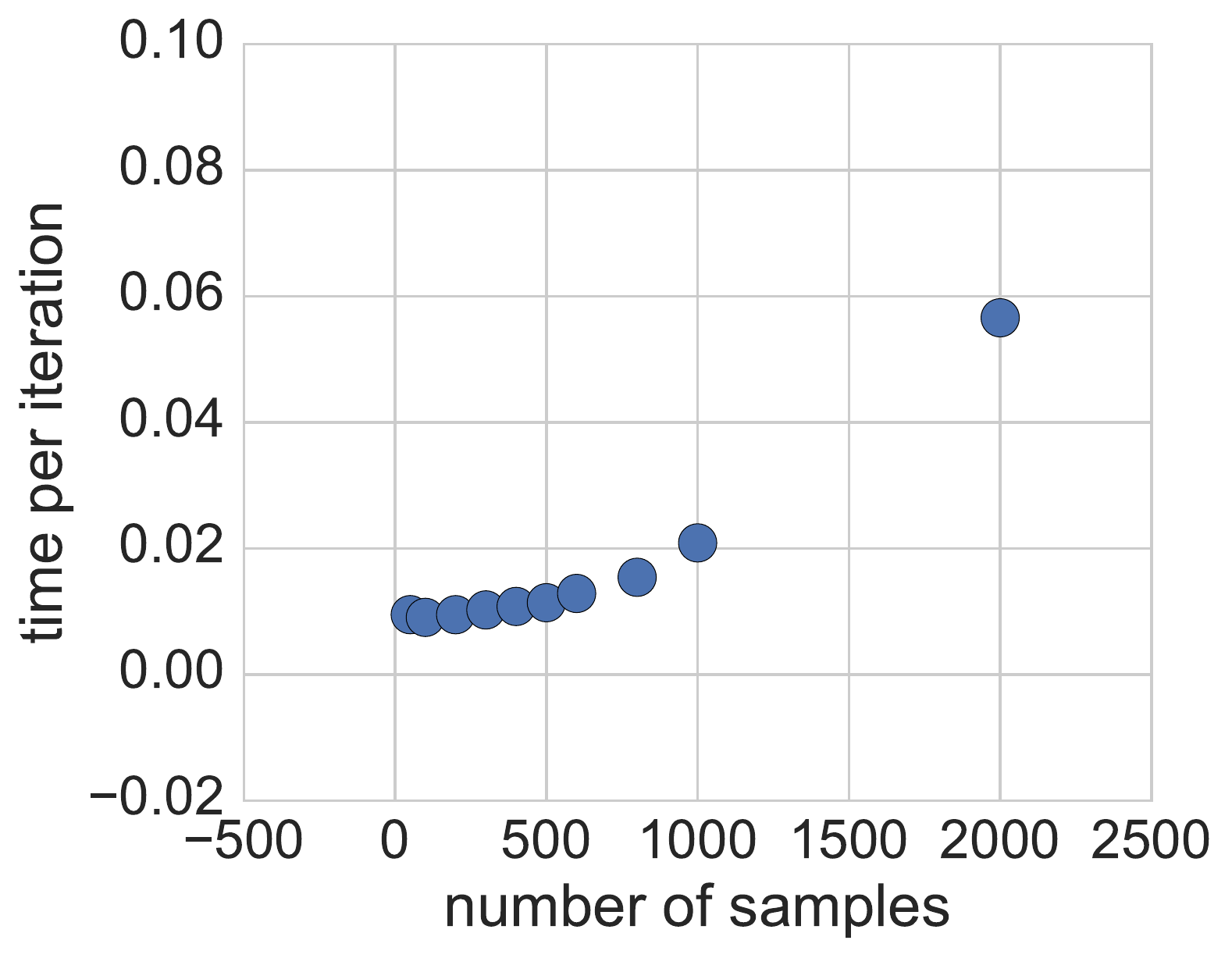}
    \caption{Time (s) per iteration, as the number of samples increases.}
    \label{fig:complexity}
\end{figure}

Fig.~\ref{fig:results_conv} shows the average number of convergence steps for different values of sensitivity, specificity and sample sizes.
Interestingly, we note that for high values of the sensitivity/specificity, the rate of convergence is much faster. To illustrate the complexity of the algorithm, we also display in Fig.~\ref{fig:complexity} the time required per iteration as a function of the number of samples.
% \begin{figure}
%     \centering
%     \includegraphics[width=0.6\textwidth]{FIG_SIM/convergence_alg5.pdf}
%     \caption{Number of steps until convergence for $n=300$, $\sigma = 0.5$ and different values of sensitivity (simulated data).}
%     \label{fig:results_conv}
% \end{figure}

\section{Additional Plots on Real Data}

\begin{figure}
    \centering
    \includegraphics[width=\textwidth]{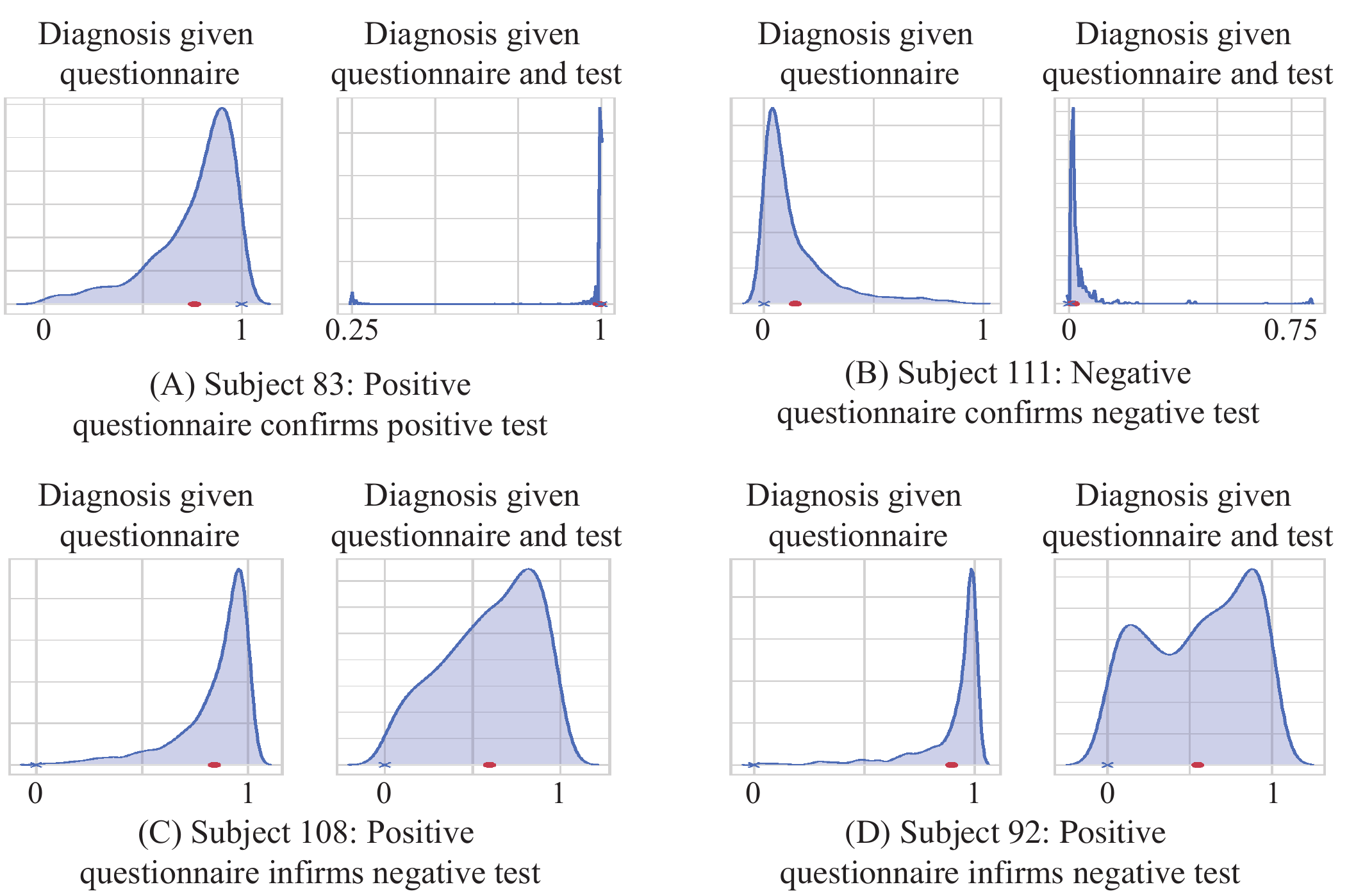}
    \caption{Posterior diagnosis distributions on four selected subjects. In each panel (A-D): the left plot represents the posterior of the diagnosis, given the symptoms and risk factors data reported in the questionnaire, while the right plot is the posterior of the diagnosis, given the symptoms, risk factors, and reported result of the diagnostic test. The red dot represents the expectation of the corresponding distribution.}
    \label{fig:examples}
\end{figure}

\xhdr{At the subject level} Fig.~\ref{fig:examples} presents four examples, where our algorithm either confirms or infirms the result of the test --- thereby allowing for the potential flagging of false negatives or negatives. 
%h specificity of the test (99\%), the probability of encountering cases where we would want to infirm positive tests is extremely small.

The first example (Fig.~\ref{fig:examples} A) is a user that registers a positive test, together with a significant number of symptoms and risk factors. The second user (Fig.~\ref{fig:examples} B) registers a negative test, while being asymptomatic and with a limited number of risk factors. In both cases, we expect our model to confirm the result of the test, and provide a narrower confidence interval as per the probability of immunity --- as confirmed by Fig.~\ref{fig:examples}.

The third and fourth examples showcase instances where the our diagnostic posterior and the test disagree. Subject 108 is a user that registers a negative test, while exhibiting a wide number of known COVID symptoms (dry cough, shortness of breath, fever), but took the test less than 10 days after his illness. Similarly, subject 92 exhibited less symptoms (in particular, no shortness of breath), but lives in a household of three, where all the other members have also fallen sick. While the posteriors reclassify the subjects' diagnosis in each case, the confidence interval associated with the prediction of immunity reflect the uncertainty that is associated with these cases, and flag them as potential false negatives.

\begin{figure}
\centering
    \includegraphics[width=\textwidth]{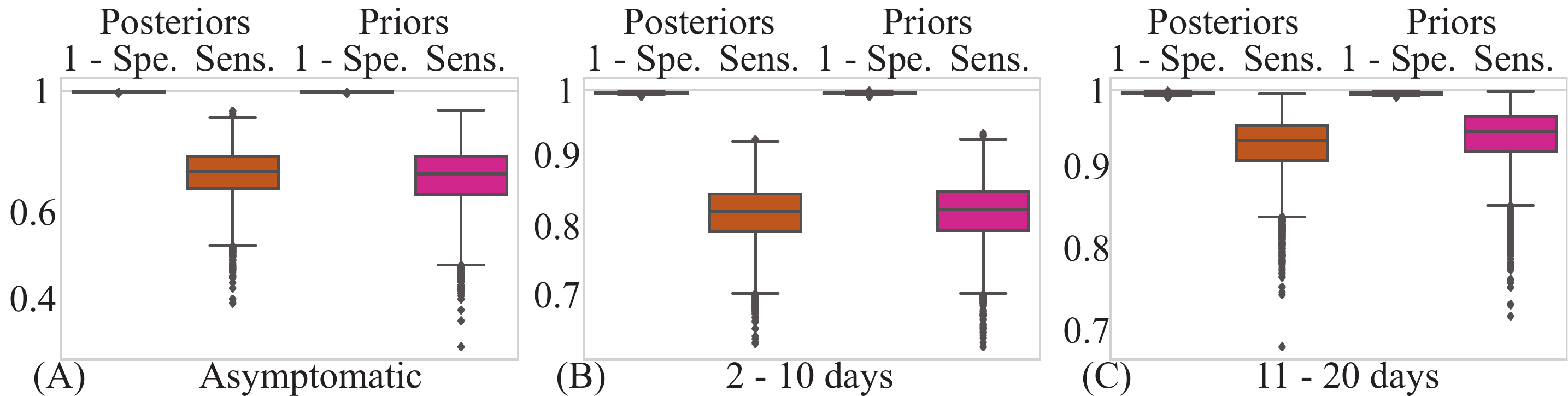}
    \caption{Posteriors of sensitivity and specificity compared to their priors, for three regimes: (A) Asymptomatic, (B) LFA test taken 2-10 days after first symptoms, (C) LFA test taken 11-20 days after first symptoms.}
    \label{fig:real_spe_sens}
\end{figure}

\xhdr{At the global level: LFA sensitivity and specificity} The posterior distributions of the model's parameters shed light on the actual accuracy of the LFA tests on our population. Fig.~\ref{fig:real_spe_sens} compares the posteriors of the sensitivity and specificity to the priors built from values reported by manufacturers, for: (A) asymptomatic subjects, (B) symptomatic subjects who took the test between 2 to 10 days after their first symptoms, and (C) symptomatic subjects who took the test between 11 to 20 days after their first symptoms. We observe that the StEM has updated the prior distributions of sensitivity and specificity, the most significant update being observed for the asymptomatic subjects.

\begin{figure}
    \centering
    \includegraphics[width=\textwidth]{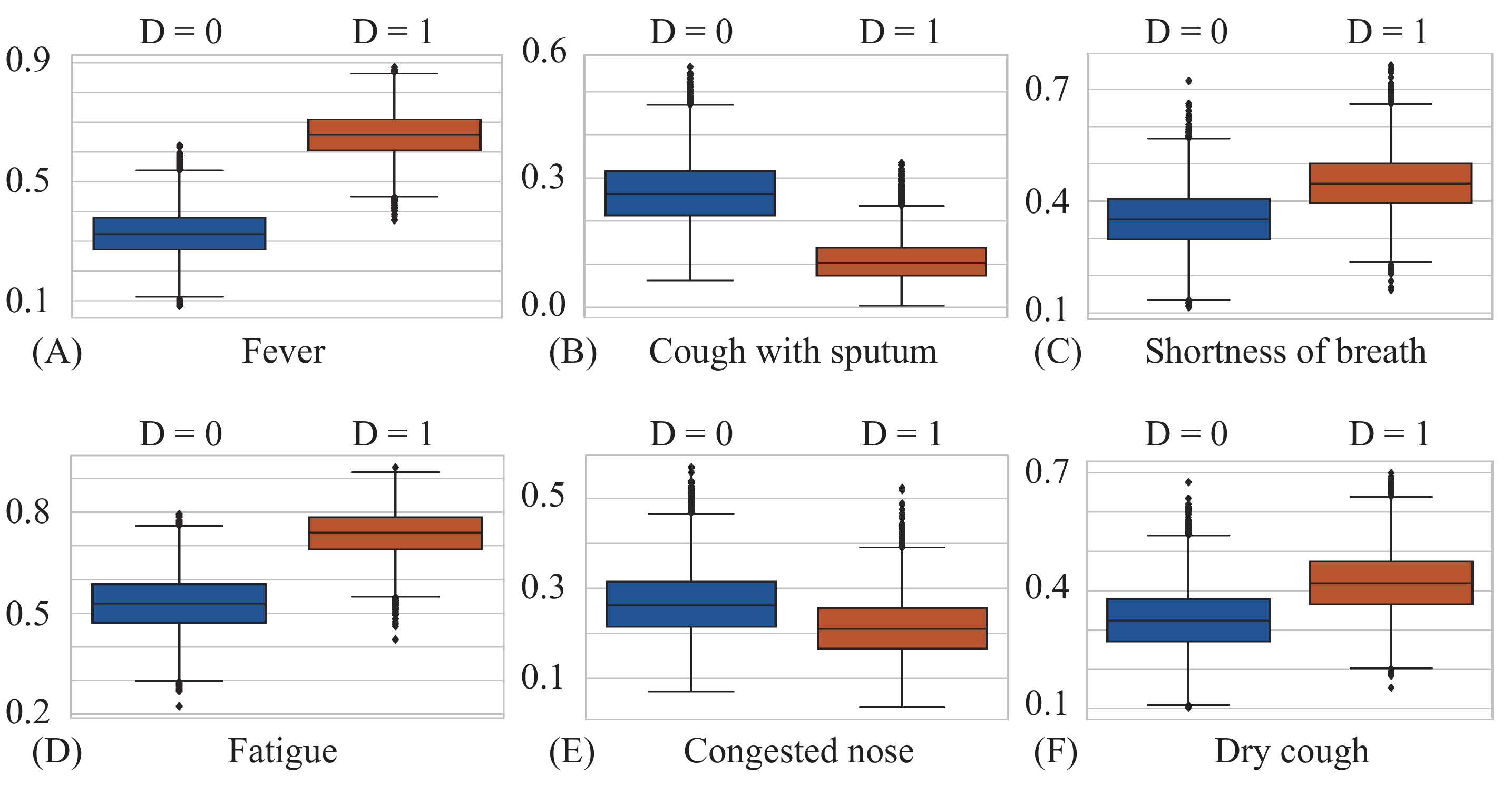}
    \caption{Posterior distributions of the probability of exhibiting selected symptoms, for a symptomatic subject. The probability of exhibiting each symptom (A-F) is plotted for symptomatic subjects with estimated negative ($D=0$) or positive ($D=1$) diagnosis.}
    \label{fig:real_posterior}
\end{figure}

\xhdr{At the global level: COVID-19 symptoms in the population of interest} Furthermore, our results provide information regarding the most prevalent symptoms for COVID-19. The posterior probability of exhibiting specific symptoms, among symptomatic subjects with positive or negative predicted diagnostic is shown on Fig.~\ref{fig:real_posterior}. We emphasize that these probability distributions are relevant to our population of $n=117$ healthcare workers, and may vary for studies considering another population.

\end{document}